\documentclass[aps, physrev, reprint, superscriptaddress, longbibliography, floatfix]{revtex4-2}

\usepackage[utf8]{inputenc}
\usepackage{amsmath}
\allowdisplaybreaks
\usepackage{amssymb}
\usepackage{hyperref}
\usepackage{graphicx}
\usepackage{dcolumn}
\usepackage{bm}
\usepackage{soul}
\usepackage{xspace}
\usepackage{verbatim}
\usepackage{color}
\usepackage{booktabs}  
\usepackage{siunitx}   
\usepackage{subcaption}

\newcommand{\bra}[1]{\langle #1 |}
\newcommand{\ket}[1]{| #1 \rangle}

\hypersetup{colorlinks=true,linkcolor=blue,citecolor=blue, filecolor=blue,urlcolor=blue,breaklinks=true}

\begin{document}

\title{Variational simulation of higher-spin systems on qubit-based quantum simulators}

\newcommand{\UESTC}{Institute of Fundamental and Frontier Sciences, University of Electronic Science and Technology of China, Chengdu 611731, China}
\newcommand{\QPPQI}{Key Laboratory of Quantum Physics and Photonic Quantum Information, Ministry of Education, University of Electronic Science and Technology of China, Chengdu 611731, China}
\newcommand{\HUAWEI}{Central Research Institute, Huawei Technologies, Shenzhen 518129, China}
\newcommand{\SIQSE}{Shenzhen Institute for Quantum Science and Engineering, Southern University of Science and Technology, Shenzhen 518055, China}
\newcommand{\IQA}{International Quantum Academy, Shenzhen 518048, China}
\newcommand{\GPKLQSE}{Guangdong Provincial Key Laboratory of Quantum Science and Engineering, Southern University of Science and Technology, Shenzhen 518055, China}
\newcommand{\SKLQSE}{Shenzhen Key Laboratory of Quantum Science and Engineering, Southern University of Science and Technology, Shenzhen 518055, China}

\author{Chufan Lyu}
\email{chufan.lyu1@std.uestc.edu.cn}
\affiliation{\UESTC}
\affiliation{\QPPQI}

\author{Zuoheng Zou}
\affiliation{\HUAWEI}

\author{Xusheng Xu}
\affiliation{\HUAWEI}

\author{Man-Hong Yung}
\affiliation{\HUAWEI}
\affiliation{\SIQSE}
\affiliation{\IQA}
\affiliation{\GPKLQSE}
\affiliation{\SKLQSE}

\author{Abolfazl Bayat}
\email{abolfazl.bayat@uestc.edu.cn}
\affiliation{\UESTC}
\affiliation{\QPPQI}

\begin{abstract}

Qubit-based quantum simulators naturally target two-level systems, whereas many quantum many-body problems are intrinsically \(d\) level. Encodings from qudits to qubits then enlarge the Hilbert space and can introduce unphysical states that interfere with variational optimization. We formulate a variational framework for encoded \(d\)-level models that suppresses these illegitimate states with penalty terms and benchmark it for spin-1 and spin-3/2 bilinear-biquadratic Heisenberg chains. We compare binary encoding, which minimizes the qubit overhead, with symmetry encoding, which preserves the relevant spin symmetries and enables symmetry-conserving ansätze. Although binary encoding is more qubit efficient, its hardware-efficient ansatz is harder to train and less effective at exploiting conserved quantities. In contrast, symmetry encoding requires more qubits but reaches substantially higher fidelities, converges faster, and exhibits better trainability than the binary hardware-efficient ansatz. These results identify symmetry-preserving encodings as a practical route to simulating higher-spin models on existing qubit platforms.

\end{abstract}

\maketitle

\section{Introduction} \label{sec:intro}

The importance of quantum computation lies in their capability for exponential speedup in simulating quantum systems as well as addressing certain computational problems. Quantum computers are programmable machines that can implement any unitary operation on quantum systems and thus convert any quantum state in the Hilbert space to another. Any unitary operation can be decomposed into a combination of single-qubit rotations as well as a two-qubit entangling gate, the so-called universal gates. This creates the concept of digital quantum computation in which the design of a quantum computer reduces to implementation of the universal gates. Near-term noisy intermediate scale quantum (NISQ) devices, however, suffer from various imperfection, including limited connectivity, short coherence time, noisy gate operations, absence of quantum error correction, imperfect initialization, and faulty readout~\cite{Preskill2018quantumcomputingin}. These imperfections make universal quantum computation inaccessible, at least, in a near future for NISQ devices. Nonetheless, quantum simulators, which are programmable machines with the capability of realizing a limited class of unitary operations for emulating complex quantum systems, are feasible~\cite{Feynman1982,lloyd2018quantum}. Thanks to recent advancement in quantum technologies, quantum simulators are emerging in various physical platforms, including optical lattices~\cite{bordia2017probing,schreiber2015observation,gross2017quantum}, Rydberg atoms~\cite{omran2019generation,Keesling2019}, optical chips~\cite{wang2017experimental,zhong2020quantum,carolan2020variational}, solid-state systems~\cite{li2017measuring}, ion-traps~\cite{lanyon2011universal,zhang2017observation,hempel2018quantum,kokail2019self}, and superconducting devises~\cite{Han2024,salathe2015digital,wang2020efficient,karamlou2021analyzing,neill2021accurately,han2021experimental,braumuller2022probing,Zhang2022Digital,shi2022observing}. These implementations are mostly qubit based, making them naturally suitable for simulating two-level quantum systems.

Many systems in nature are inherently $d$ level (with $d>2$), including higher spins~\cite{levitt2013spin}, bosons~\cite{Fisher1989}, vibrational modes~\cite{Wilson1955}, and itinerant electrons~\cite{M_Shimizu_1981}.
In addition, there are classical optimization problems that can efficiently mapped to qudit systems~\cite{Glos2022, tabi2020quantum}.
In principle, there are two methods for simulating such systems on quantum simulators. First, one can design and build $d$-level quantum simulators that can directly emulate those systems~\cite{Liu2023,meth2024simulating,Zache2023fermionquditquantum,Daniel2022,Hrmo2023,Ringbauer2022,Luo2014}. Second, one can rely on existing qubit-based quantum simulators but exploit proper encoding to effectively reproduce the behavior of $d$-level systems. The former approach is still technologically challenging, in particular, for realizing a set of universal gate operations~\cite{Liu2023,meth2024simulating,Zache2023fermionquditquantum,Daniel2022,Hrmo2023,Ringbauer2022,Luo2014}. The latter approach requires suitable encoding methods to map $d$-level systems to qubits.
In Refs.~\cite{Wu2002,Batista2004,Sawaya2020,Dicke1954, yang2024cost, Chen2021, Sawaya2023, Berwald2022, Karimi2019, Chancellor2019, DiMatteo2021, Tilly2022}, several encoding methods have been introduced to map a \(d\)-level system onto a qubit system.
In these methods, depending on the mapping algorithm, a single $d$-level system can be mapped to a different number of qubits. While the most efficient mapping requires \(\lceil \log_2(d) \rceil\) qubits, 
the other may demand $d$ qubits. Note that the complexity does not solely depend on the mapping as the manipulation of the qubits might be easier for a less efficient encoding.
Simulating \(d\)-level systems on qubit-based quantum simulators may face several challenges: (1) Mapping \(d\)-level systems to qubit systems can enlarge the corresponding Hilbert space, implying that certain regions of the Hilbert space does not represent the original system; (2) the resulting qubit Hamiltonian may include multibody interactions, which complicates the quantum simulation; and (3) symmetries of the original Hamiltonian may take complex forms after mapping to the qubits.

While universal quantum computation is not achievable with NISQ simulators, one may wonder whether such quantum devices can provide any computational advantage over classical computers. Indeed, the advantage of NISQ simulators has been demonstrated in sampling problems~\cite{zhong2020quantum,Zhong2021Phase,Deng2023Gaussian}. However, sampling problems do not have practical applications. Hence, achieving practical advantage with NISQ simulators is still an open problem. Variational quantum algorithms (VQAs) are the most promising approach for achieving quantum advantage in NISQ era. Variational quantum algorithms have been developed for a range of applications, including quantum chemistry~\cite{google2020hartree,o2016scalable,kandala2017hardware,hempel2018quantum,colless2018computation,nam2020ground}, combinatorial optimization~\cite{farhi2014quantum,bravyi2020obstacles,Zimboras2024,Glos2022}, quantum machine learning~\cite{biamonte2017quantum,arunachalam2017survey,ciliberto2018quantum,dunjko2018machine,farhi2018classification,schuld2019quantum,wilkinson2022evaluating,zapletal2023errortolerant}, dynamical simulations in closed~\cite{cirstoiu2020variational,gibbs2021longtime,yuan2019theory,mcardle2019variational,heya2019subspace} and open~\cite{huh2014linear,hu2020quantum,endo2020variational,yuan2019theory,haug2020generalized} systems, quantum sensing~\cite{meyer2021variational,meyer2021fisher,beckey2020variational,kaubruegger2019variational,koczor2020variational,ma2021adaptive,haug2021natural}, and condensed matter physics~\cite{kokail2019self,sagastizabal2021variational,Lyu2020accelerated}.
Variational quantum eigensolver (VQE) is an important class of VQAs that tend to simulate the low-energy eigenstates of a given Hamiltonian~\cite{kandala2017hardware,nakanishi2019subspace,higgott2019variational,mcclean2017hybrid,santagati2018witnessing,hong2023quantum,Verena2022}. Exploiting the inherent symmetries of the Hamiltonian in the VQE protocol, either through tailored circuit design~\cite{barkoutsos2018quantum,wang2009efficient,Lyu2020accelerated,seki2020symmetry,Gard2020,barron2021preserving,zhang2021shallow,zheng2021speeding,anselmettiLocalExpressiveQuantumnumberpreserving2021} or by incorporating them into the cost function as penalty terms~\cite{mcclean2016theory,ryabinkin2018constrained}, is extremely beneficial in simplifying the protocol. However, most of these symmetry exploiting methods have been focused on qubit systems. While VQE has also been applied to both fermionic and bosonic systems~\cite{anselmettiLocalExpressiveQuantumnumberpreserving2021,Dutta2025bosonic,Nigmatullin2025compact,Ilyas2022ternary,Zache2023fermionquditquantum,Li2023fermionic}, a systematic symmetry investigation of $d$-level quantum systems remains unexplored.

In this work, we study variational simulations of $d$-level systems on qubit-based quantum simulators. We first show that mapping a $d$-level model to qubits can introduce illegitimate states in the enlarged Hilbert space, and we present a systematic penalty construction that suppresses those states during optimization. We then compare binary and symmetry encoding methods. Binary encoding is qubit efficient, but it does not naturally incorporate Hamiltonian symmetries in the circuit design. Symmetry encoding requires additional qubits, but it preserves the relevant symmetries and enables symmetry-conserving variational circuits. Through extensive ground-state simulations for higher-spin chains, we show that symmetry encoding yields higher fidelities, faster convergence, and better trainability than the binary hardware-efficient ansatz. We also determine the required quantum resources for specific models at different points of their phase diagram. The framework is directly compatible with existing qubit-based platforms and extends to a broad class of qudit models.

\section{Variational framework for encoded qudit simulation}

\subsection{Variational quantum eigensolver and resource metrics}

Variational quantum algorithms have emerged as a promising class of algorithms that leverage the hybrid quantum-classical approach to solve optimization problems and simulate quantum systems, particularly beneficial in the NISQ era. In this section, we provide a brief introduction to the VQAs, focusing on their applications in preparing the ground state of a given Hamiltonian or simulating nonequilibrium dynamics using the shallowest possible quantum circuit.

The VQE is a pivotal example of VQA dedicated to finding the low-energy eigenstates (e.g., the ground state) of a Hamiltonian, which is central to understanding the physical properties of quantum systems. In the VQE algorithm, a quantum state, denoted as $\ket{\psi(\vec{\theta})}$, is prepared through a parametrized quantum circuit acting on an initial state, expressed as $\ket{\psi(\vec{\theta})}{=}U(\vec{\theta})\ket{\psi_0}$, for a given $N_q$ qubits.
This parametrized quantum circuit, often referred as the ansatz, incorporates a set of tunable parameters $\vec{\theta}=(\theta_1,\theta_2,\dots,\theta_L)$, and $\ket{\psi_0}$ is the initial state.
Varying $\vec{\theta}$ enables the output of the quantum circuit, namely, $\ket{\psi(\vec{\theta})}$, to explore a segment of the Hilbert space, which is called reachable set. Notably, in circuits of significant depth, and thus a large number of parameters $L$, the reachable set can span the entire Hilbert space, allowing the preparation of any quantum state of $N_q$ qubits. However, the goal is to keep the circuit as shallow as possible, focusing on the most relevant part of the Hilbert space. Specifically, VQE aims for a comparatively shallow ansatz that the reachable set contains the ground state of the Hamiltonian of interest $H$. Various ansätze have been proposed, each differing in complexity~\cite{kandala2017hardware,Lyu2023symmetryenhanced,Wiersema2020,Haegeman2012,Vicentini2019,huang2022robust}.
The expectation value of the Hamiltonian $\langle H \rangle = \bra{\psi(\vec{\theta})}H\ket{\psi(\vec{\theta})}$ is measured at the output of the simulator. One then iteratively minimizes $\langle H \rangle$ with respect to the parameters $\vec{\theta}$ using a classical optimization algorithm, such as Adam~\cite{Kingma2014adam}. Ideally, after a few iterations, $\langle H \rangle$ converges to its minimum for a special set of parameters $\vec{\theta}^*$. If the reachable set contains the ground state, then the variational circuit with optimal parameters $\vec{\theta}^*$ produces the ground state of the Hamiltonian $H$.

In the domain of VQAs, resource efficiency emerges as a critical property due to the constraints imposed by NISQ devices. These devices are typically characterized by a limited number of qubits, brief coherence times, and elevated error rates. As a result, a quantification of resource efficiency is essential for assessing the performance of VQAs. This efficiency can be dissected into two primary categories: classical and quantum resources.
Classical resources primarily depend on the computational demands of the optimization process, notably, the number of iterations needed to tune the parameters. Furthermore, each parameter requires individual measurements for gradient evaluation if the optimizer in the VQAs adopted to gradient-based optimizer. Thus, we define the measure of classical resources ($C_R$) as the product of the number of parameters $L$ and the number of optimization iterations $n_I$, expressed mathematically as
\begin{equation}\label{eq:classical_resources}
C_R = L \times n_I.
\end{equation}
For quantum resources, the emphasis is on the circuit components that are most susceptible to noises on NISQ devices. Since single-qubit operations are relatively error resilient and precise, the focus shifts to two-qubit entangling operations, which are more prone to errors. Hence, we quantify quantum resources by the number of two-qubit entangling gates used in the circuit.

\begin{figure*}[ht]
  \centering
  \begin{subfigure}[b]{0.95\columnwidth}
    \includegraphics[width=\linewidth]{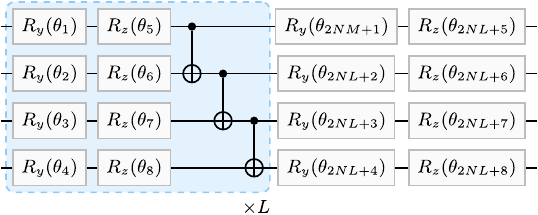}
    \caption{}
    \label{fig:hardware_eff}
  \end{subfigure}
  \hspace{2em}
  \begin{subfigure}[b]{0.75\columnwidth}
    \includegraphics[width=\linewidth]{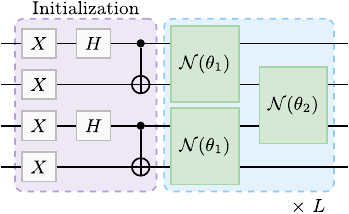}
    \caption{}
    \label{fig:stot_ansatz}
  \end{subfigure}
  
  \vspace{1em}
  
  \begin{subfigure}[b]{0.95\columnwidth}
    \includegraphics[width=\linewidth]{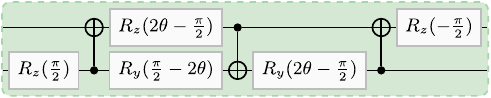}
    \caption{}    
    \label{fig:N_func}
  \end{subfigure} 
  \hspace{2em}
  \begin{subfigure}[b]{0.75\columnwidth}
    \centering
    \includegraphics[width=0.7\linewidth]{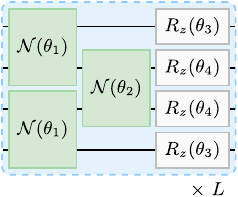}
    \caption{}    
    \label{fig:sz_ansatz}
  \end{subfigure}
  
  \caption{Variational ansatz circuits used in this work. 
  (a) The hardware-efficient ansatz circuit for system size $N_q{=}4$. The circuits highlighted in blue can be repeated to increase the expressibility of the variational circuit. 
  (b) The total spin conserving ansatz circuit for system size $N_q{=}4$.
  (c) Circuit implementing of $\mathcal{N}(\theta_x, \theta_y, \theta_z)$ as an entangling gate between two qubits. 
  (d) The $s_z$-conserving ansatz circuit for system size $N_q{=}4$. }
  \label{fig:ansatzes}
\end{figure*}

Therefore, in consideration of developing a VQA, selecting an appropriate ansatz is essential for achieving an optimal balance between minimizing the use of error-prone two-qubit gates and controlling the number of variational parameters. The former is crucial for quantum resource efficiency, and the latter not only dictates the expansiveness of the reachable set but also influences the efficiency of classical resources. 

One of the most commonly used ansätze in the literature is the hardware-efficient ansatz, depicted in Fig.~\ref{fig:ansatzes}(a). This ansatz consists of alternating layers of single-qubit rotations and two-qubit entangling gates. The single-qubit rotations are defined by the formula
\begin{equation}
  R_\alpha(\theta) = e^{-i\theta \sigma^\alpha/2},
\end{equation}
where $\sigma^\alpha$ is the Pauli operator, with $\alpha \in \{x, y, z\}$. For the two-qubit operations, we employ the Controlled-not (CNOT) gate, which is specified by the matrix
\begin{equation}
  \text{CNOT} = \begin{pmatrix} 1 & 0 & 0 & 0 \\ 0 & 1 & 0 & 0 \\ 0 & 0 & 0 & 1 \\ 0 & 0 & 1 & 0 \end{pmatrix}.
\end{equation}
This structure is particularly favored for its simplicity and efficiency in creating entanglement between qubits, which is critical for the success of various variational quantum algorithms. 
Nonetheless, the hardware-efficient ansatz suffers from barren plateau~\cite{McClean2018,Holmes2022}, which makes its training challenging for large-scale quantum systems. Furthermore, this ansatz does not inherently preserve any symmetries that may exist in the Hamiltonian, which could enhance quantum simulation performance. As an alternative, one might consider a more complex entangling gate with tunable parameters, such as~\cite{Vatan2004}
\begin{equation}
  \mathcal{N}(\theta_x, \theta_y, \theta_z) = e^{i (\theta_x \sigma^x \otimes \sigma^x + \theta_y \sigma^y \otimes \sigma^y + \theta_z \sigma^z \otimes \sigma^z)},
\end{equation}
where $\sigma^x \otimes \sigma^x$, $\sigma^y \otimes \sigma^y$, and $\sigma^z \otimes \sigma^z$ are the tensor products of the Pauli operators acting on two qubits. The circuit implementation of this unitary operation is depicted in Fig.~\ref{fig:ansatzes}(c). In the special case where \(\theta_x = \theta_y = \theta_z = \theta\), this unitary conserves the number of excitations as well as the total spin. By combining this entangling gate with single-qubit rotations, for example, $R_z(\theta)$, one can construct an excitation preserving ansatz, $U(\vec{\theta})$, satisfying $[U(\vec{\theta}), S^z_{tot}]=0$, where $S^\alpha_{tot} = \frac{1}{2}\sum_i \sigma^\alpha_{i}$ for \(\alpha \in \{x, y, z\}\), as illustrated in Fig.~\ref{fig:ansatzes}(d). In the absence of single-qubit rotations on the $z$ axis, namely, $R_z(\theta)$, the ansatz preserves the total spin, i.e., $[U(\vec{\theta}), S_{\text{tot}}^2]=0$, where $S_{\text{tot}}^2 = {(S_{tot}^x)}^2 + {(S_{tot}^y)}^2 + {(S_{tot}^z)}^2$, shown in Fig.~\ref{fig:ansatzes}(b).

\subsection{Binary and symmetry-preserving encodings}

The first step toward the simulation of a $d$-level many-body quantum system on a qubit-based quantum simulator is to exploit an encoding algorithm that maps the $d$-level operators into qubit operators. Several encoding techniques have been explored~\cite{Wu2002,Batista2004,Sawaya2020,Dicke1954}. The most straightforward method is the binary encoding in which a $d$-level particle is mapped to $M{=}\lceil \log_2(d) \rceil$ qubits, where $\lceil x \rceil$ represents the smallest integer number that is greater or equal to the real number $x$. In this encoding, each $d$-level basis state $\ket{\widetilde{j}}$ ($j=0,1,\ldots,d-1$) is mapped to the binary representation of its index $j$, which is converted into an $M$-qubit computational state. The binary representation is padded with zeros on the left to match the number of qubits $M$:
\begin{equation}\label{eq:binary_mapping}
\ket{\widetilde{j}} \mapsto \ket{\mathbf{b}_M^{(j)}}=\ket{b_{M-1} b_{M-2} \cdots b_1 b_0},
\end{equation}
where $b_{M-1} b_{M-2} \cdots b_1 b_0$ is the binary representation of the integer $j$ padded to $M$ bits. 
For instance, for a nine-level qudit system, we require four qubits to encode $\ket{\widetilde{0}}=\ket{0000}$ all the way to $\ket{\widetilde{8}}=\ket{1000}$. Note that throughout this paper, the tilde basis $\ket{\widetilde{j}}$ is used for qudits and the ordinary basis $\ket{j}$ is used for qubits.
It will be convenient to define the transformation operator between the two bases as
\begin{equation}\label{eq:binary_transformation}
    \mathcal{T}_b=\sum_{j=0}^{d-1} \ket{\mathbf{b}_M^{(j)}} \bra{\widetilde{j}}.
\end{equation}
Note that while we always have $\mathcal{T}_b^\dagger \mathcal{T}_b=\mathbb{I}_{\mathrm{qudits}}$, the vice versa is not always valid, i.e., $\mathcal{T}_b \mathcal{T}_b^\dagger \neq \mathbb{I}_{\mathrm{qubits}}$, unless $M{=}\log_2(d)$. By using the transformation matrix $\mathcal{T}_b$, one can map any qudit operator
$\widetilde{H}$ into a qubit basis through $H=\mathcal{T}_b \widetilde{H} \mathcal{T}_b^\dagger$ and then write it in terms of Pauli strings with the help of the following maps:
\begin{eqnarray}\label{eq:op_mapping}
  \ket{0}\bra{0} &=& \frac{1}{2}(\mathbb{I} + \sigma^z), \;\;\;\;\;\;
  \ket{1}\bra{1} = \frac{1}{2}(\mathbb{I} - \sigma^z),  \nonumber \\
  \ket{0}\bra{1} &=& \frac{1}{2}(\sigma^x + i\sigma^y), \;\;  \ket{1}\bra{0} = \frac{1}{2}(\sigma^x - i\sigma^y),
\end{eqnarray}
where $\mathbb{I}$ is the identity and $\sigma^{x,y,z}$ are the Pauli operators.

If the qudit Hamiltonian $\widetilde{H}$ supports a symmetry $\widetilde{S}$, namely, $[\widetilde{H},\widetilde{S}]{=}0$, the corresponding qubit Hamiltonian $H=\mathcal{T}_b\widetilde{H}\mathcal{T}_b^\dagger$ supports an equivalent symmetry $S=\mathcal{T}_b\widetilde{S}\mathcal{T}_b^\dagger$. However, even if $\widetilde{S}$ might be a simple local operator the form of $S$ might be complex consisting of several multiqubit terms in the qubit basis. It has been shown that in VQE simulations, the most resource efficient approach to benefit from the symmetries is to incorporate them in the design of the circuit such that $[U(\vec{\theta}), S]{=}0$~\cite{Lyu2023symmetryenhanced}. In this situation, the circuit preserves the symmetries in the initial state and thus makes the VQE simulation easier. The problem is that if the symmetry operator $S$ is complex itself then designing such a circuit becomes challenging. This is often the case for mapping qudits to qubits. Therefore, an alternative mapping might be exploited, namely, symmetry encoding, which is also known as Dicke state encoding~\cite{Dicke1954}.
This approach maps each \(d\)-level basis state \(\ket{\widetilde{j}}\) into a superposition of \(\binom{d-1}{j}\) computational qubit states with exactly \(j\) excitations: 
\begin{equation}
\ket{\widetilde{j}} \mapsto \ket{\mathbf{f}^{(j)}_{d-1}}= \frac{1}{\sqrt{\binom{d-1}{j}}} \sum_{\substack{ \forall k \text{ with } j\\ \text{excitations}  }} \ket{\mathbf{b}_{d-1}^{(k)}},
\end{equation}
where $\binom{d-1}{i}$ is the binomial coefficient that guarantees the normalization, $\ket{\mathbf{b}_{d-1}^{(k)}}$ is given in Eq.~(\ref{eq:binary_mapping}) and summation runs over all values of $k=0,\ldots,d-1$ in which $\ket{\mathbf{b}_{d-1}^{(k)}}$ has only $j$ excitations. One can define the transformation operator between the two bases as
\begin{equation}
  \mathcal{T}_s=\sum_{j=0}^{d-1} \ket{\mathbf{f}_{d-1}^{(j)}} \bra{\widetilde{j}}.
\end{equation}

The symmetry encoding scheme effectively preserves certain types of symmetries, for example, the conservation of components of the total spin in a spin system, i.e., $[\widetilde{H}, \widetilde{S}^{\alpha}_{tot}] = [H, S^{\alpha}_{tot}] = 0$, where $\widetilde{S}^\alpha_{tot} = \sum_i \widetilde{S}^{\alpha}_{i}, (\alpha{=}x,y,z)$ (similar definition is given for $S^{\alpha}_{tot}$ in the qubit basis). Here, $\widetilde{S}^{\alpha}_{i}$ is the $\alpha$ component of the spin operator acting on the $i$th site of the $d$-level system.
This conservation implies that the symmetry-preserving encoding also preserves the conservation of total spin from $d$ level to qubit systems, i.e., $[\widetilde{H}, {\widetilde{S}_{\text{tot}}}^2] = [H, S_{\text{tot}}^2] = 0$, where $\widetilde{S}_{\text{tot}}^2 = (\widetilde{S}_{tot}^{x})^2 + (\widetilde{S}_{tot}^y)^2 + (\widetilde{S}_{tot}^z)^2$ (again similar definition is valid for $S_{\text{tot}}^2$ in the qubit basis).

Although, in general, using the symmetry encoding demands extra qubits than the binary one, it allows one to use ansätze that have been developed for preserving certain symmetries such as total spin or number of excitations~\cite{Gard2020,Lyu2023symmetryenhanced}. Such ansätze can improve trainability and reduce the classical optimization cost compared to a conventional hardware-efficient ansatz that does not preserve symmetries. But this advantage does not necessarily imply a smaller two-qubit gate count. Therefore, the final performance of symmetry encoding compared to binary encoding is determined by a trade-off between the additional qubits and possible circuit overhead on the one hand, and the improved fidelity and trainability enabled by symmetry-preserving circuits on the other. Determining which of these factors dominates in variational simulations of $d$-level systems is the subject of this paper.

It is important to note that, for both of the encoding methods, the Hilbert space of the qubit system is larger than (or at least equal to) the one for the qudit system. For instance, in a many-body system of $N_d$ qudit particles, the size of the Hilbert space is $d^{N_d}$. While using the binary encoding enlarges the size of the Hilbert space to $2^{N_d M}$, using symmetry encoding enlarges it even more to $2^{N_d(d-1)}$. As a result, certain states within the encoded qubit system are ``illegitimate,'' i.e., having no counterpart in the qudit system. Thus, for any quantum state $\ket{\chi}$ within the illegitimate subspace, one inherently gets $H\ket{\chi}=0$.
The illegitimate states can indeed complicate the simulation process, particularly in VQE implementations, where such states may erroneously appear as the state with minimum energy. This issue typically arises when the actual ground state of the \(d\)-level system has an energy expectation value greater than zero, namely, \(\langle \widetilde{gs} | \widetilde{H} | \widetilde{gs} \rangle >0\).
Under these conditions, the illegitimate states, which inherently possess zero energy, will be the ground state of the encoded qubit system. To address this issue, one can incorporate a penalty term into the cost function of the VQE that specifically pushes the energy of these states upward. 

To exemplify the inclusion of penalty term, consider a symmetry encoding on a spin-$1$ system with three-level basis states
\begin{equation}
    \ket{\widetilde{0}}=\ket{s_z{=}+1}, \quad 
    \ket{\widetilde{1}}=\ket{s_z{=}0}, \quad
    \ket{\widetilde{2}}=\ket{s_z{=}-1},
\end{equation}
where $\ket{s_z}$ represents the $z$ component of the spin-1 particles. The symmetry encoding results in
\begin{equation}\label{eq:spin1_mapping}
  \ket{\widetilde{0}} \mapsto \ket{00}, \quad \ket{\widetilde{1}} \mapsto \frac{1}{\sqrt{2}}(\ket{01} + \ket{10}), \quad \ket{\widetilde{2}} \mapsto \ket{11}.
\end{equation}
In this mapping, the singlet state $\frac{1}{\sqrt{2}}(\ket{01} - \ket{10})$ is not used and thus considered as illegitimate. To effectively penalize this state, we construct a penalty term that assigns a higher energy to the singlet state compared to the triplet states. A natural choice is the operator
\begin{equation}\label{eq:raw_penalty_term}
  \mathcal{P}_{i,i+1} = - \vec{\sigma}_i \cdot \vec{\sigma}_{i+1}.
\end{equation}
By expanding the dot product, this penalty term takes the form
\begin{align}\label{eq:spin1_penalty_term}
  \mathcal{P}_{i,i+1} = -(\sigma^x_i \otimes \sigma^x_{i+1} + \sigma^y_{i} \otimes \sigma^y_{i+1} + \sigma^z_{i} \otimes \sigma^z_{i+1}).
\end{align}
This operator has an eigenvalue of $+3$ for the singlet state and $-1$ for the triplet states. Consequently, minimizing the cost function
\begin{equation}\label{eq:penalized_cost_function}
\text{cost} = \langle H \rangle + \beta \sum_{\text{odd  } i}\langle \mathcal{P}_{i,i+1} \rangle
\end{equation}
with a sufficiently large positive scalar $\beta$ ensures that the illegitimate singlet states are energetically penalized, guiding the VQE toward the valid subspace.

{ The above encodings also determine the maximum size of the Pauli strings generated by an interaction term. For an $m$-body term in the original $d$-level model, binary encoding maps each involved $d$-level particle to $M = \lceil \log_2 d \rceil$ qubits, meaning that the longest Pauli strings act on at most $mM$ qubits. In symmetry encoding, each $d$-level particle is represented by $d - 1$ qubits, and thus the longest Pauli strings for the same $m$-body term act on at most $m(d - 1)$ qubits. Longer Pauli strings typically increase the measurement cost associated with evaluating $\langle H \rangle$. Therefore, one must balance this measurement overhead against the advantages gained by employing symmetry-preserving ansätze. Additionally, the occurrence of illegitimate states depends on both the underlying model and the chosen encoding. Once the unused basis states arising from the mapping are identified, penalty terms can be introduced to energetically suppress these states.}



\begin{figure}[ht]
  \centering
  \begin{subfigure}[b]{1\columnwidth}
    \includegraphics[width=0.95 \columnwidth]{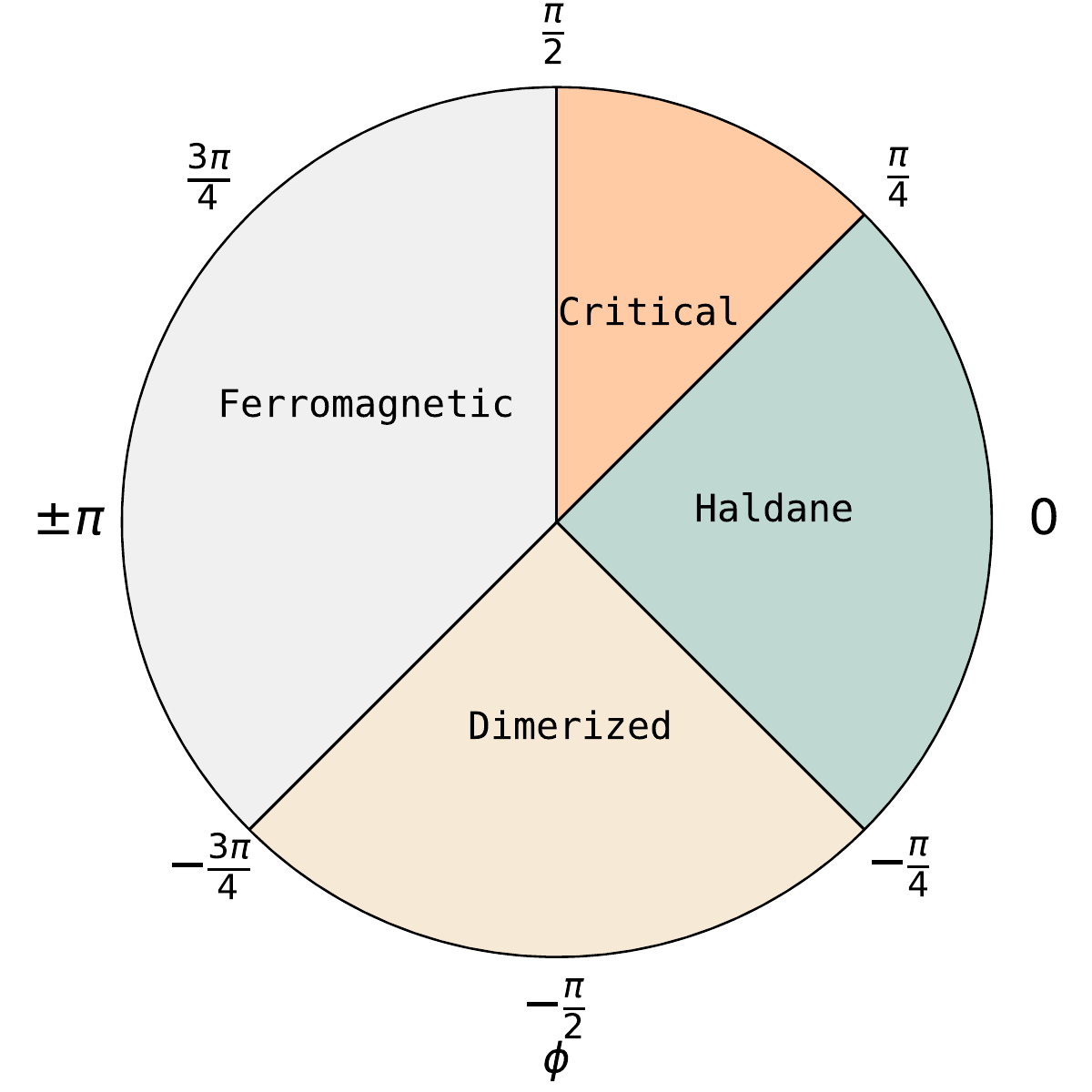}
    \caption{}
    \label{fig:phases}
  \end{subfigure} \\
  \begin{subfigure}[b]{1\columnwidth}
    \includegraphics[width=0.95 \columnwidth]{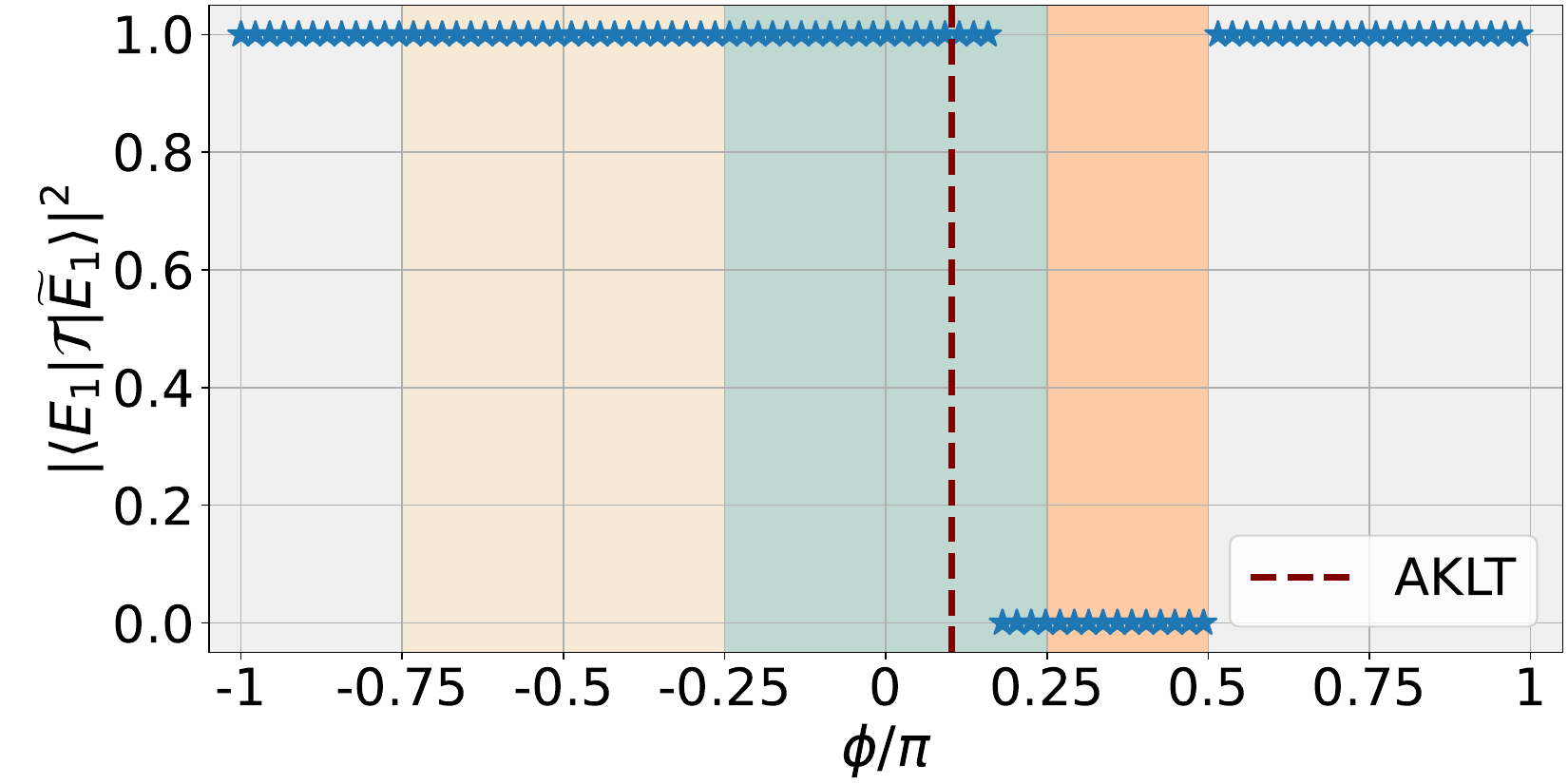}
    \caption{}
    \label{fig:check_phases}
  \end{subfigure}
  \caption{Phase diagram and ground-state fidelity validation for the spin-1 BBH model. (a) Phase diagram of the spin-1 bilinear-biquadratic Heisenberg model. (b) Fidelity between the true qudit ground state $\ket{\widetilde{E}_1}$ of the spin-1 BBH model with system size $N_d{=}6$ and the ground state of the encoded 12-qubit Hamiltonian $\ket{E_1}$ across the phase diagram, ${|\langle E_1 | \mathcal{T}| \widetilde{E}_1 \rangle |}^2$. The result is the same for both binary and symmetry encodings.}
  \label{fig:phase_diagram}
\end{figure}

\section{Spin-1 bilinear-biquadratic Heisenberg model}

Specifically, we first focus on the simulation of a spin-$1$ system (i.e., $d{=}3$). We consider the bilinear-biquadratic Heisenberg (BBH) model of size $N_d$, defined as
\begin{equation}\label{eq:BBH_spin1}
  \widetilde{H}_{\text{BBH}} = J\sum_{i=1}^{N_d-1} \left[ \cos(\phi) \vec{\widetilde{S}}_i \cdot \vec{\widetilde{S}}_{i+1} + \sin(\phi) (\vec{\widetilde{S}}_i \cdot \vec{\widetilde{S}}_{i+1})^2 \right],
\end{equation}
where $J$ is the coupling strength. In our ground-state simulation, we put $J{=}1$. $\vec{\widetilde{S}}_i = (\widetilde{S}^x_{i}, \widetilde{S}^y_{i}, \widetilde{S}^z_{i})$ represents the spin-$1$ operators at the $i$th site, with matrices given by
\begin{align}
  \widetilde{S}^x &= \frac{1}{\sqrt{2}}\begin{pmatrix} 0 & 1 & 0 \\ 1 & 0 & 1 \\ 0 & 1 & 0 \end{pmatrix}\\ \nonumber
  \widetilde{S}^y &= \frac{1}{\sqrt{2}}\begin{pmatrix} 0 & -i & 0 \\ i & 0 & -i \\ 0 & i & 0 \end{pmatrix}\\ \nonumber
  \widetilde{S}^z &= \begin{pmatrix} 1 & 0 & 0 \\ 0 & 0 & 0 \\ 0 & 0 & -1 \end{pmatrix}. 
\end{align}
The BBH model is a generalization of the Heisenberg model that includes both bilinear and biquadratic interactions between neighboring spins. The parameter $\phi$ controls the relative strength of these interactions.

{ The spin-$1$ BBH model has been widely used as a prototype for integer-spin quantum magnetism and for topological phases in one dimension. It contains the antiferromagnetic Heisenberg chain associated with the Haldane phase~\cite{Haldane1983} and the Affleck-Kennedy-Lieb-Tasaki (AKLT) point, where the valence-bond-solid ground state is exactly known~\cite{Affleck1987}. It also appears as an effective spin model for spin-$1$ bosons in optical lattices, where spin-exchange interactions can generate singlet, nematic, and dimerized phases~\cite{Imambekov2003,Chiara2011}. These applications make the model a useful benchmark for testing encoded simulations of higher-spin Hamiltonians.}

The spin-$1$ BBH model exhibits several symmetries, including the conservation of the total spin, expressed as $[\widetilde{H}_{\text{BBH}}, \widetilde{S}_{\text{tot}}^2] = 0$. This conservation law implies that every eigenstate $\ket{\widetilde{E}_k}$ of the BBH model possesses a well-defined total spin $s$, which is an integer for even system sizes $N_d$ and a half-integer for odd $N_d$. Consequently, the expectation value of the total spin operator for any eigenstate is given by $\bra{\widetilde{E}_k} \widetilde{S}_{\text{tot}}^2 \ket{\widetilde{E}_k} = s(s+1)$. Additionally, the conservation of total spin implies the conservation of its components in all directions, i.e., $[\widetilde{H}_{\text{BBH}}, \widetilde{S}^{\alpha}_{tot}] = 0$, which ensures that the $z$ component of the total spin, $\widetilde{S}_{tot}^z$, for each eigenstate is also conserved such that $\bra{\widetilde{E}_k} \widetilde{S}_{tot}^z \ket{\widetilde{E}_k} = s_z$, with $s_z$ ranging from $-s$ to $s$.

The spin-$1$ BBH model is well known in condensed matter physics for exhibiting a rich phase diagram as $\phi$ varies from $-\pi$ to $\pi$. The schematic of the phase diagram is shown in Fig.~\ref{fig:phase_diagram}(a). In the range $\phi \in \left[-\frac{3}{4}\pi, \frac{1}{2}\pi\right]$, the system is in a ferromagnetic phase, where all spins are aligned in the same direction, making the ground state trivial to simulate. We therefore focus our VQE simulations on the nontrivial phases of the BBH model.
For $\phi \in \left[-\frac{3}{4}\pi, -\frac{1}{4}\pi\right]$, the system is in the dimerized phase with broken translational invariance and singlets of neighboring spins. The ground state in this phase transition can be characterized by the order parameter~\cite{Chiara2011}:
\[
\mathcal{O}_{\text{dimer}} = |\langle H_i - H_{i+1} \rangle|,
\]
where \(H_i = \cos(\phi) \vec{\widetilde{S}}_i \cdot \vec{\widetilde{S}}_{i+1} + \sin(\phi) (\vec{\widetilde{S}}_i \cdot \vec{\widetilde{S}}_{i+1})^2\). In large systems or periodic boundary conditions, the choice of index $i$ does not matter. Nonetheless, in finite open systems, we select \(i=2\) to avoid being at the boundary.
Notably, for $\phi \in \left[-\frac{1}{4}\pi, \frac{1}{4}\pi\right]$, the system is in the Haldane phase, which is a topological phase. Particularly, at the \(\phi = \text{arctan}(1/3)\) point, known as the AKLT point, the ground state is fourfold degenerate, including a singlet state with total spin \(s=0\) and a triplet state with total spin \(s=1\). The Haldane topological phase can be characterized by a string order parameter~\cite{Chiara2011}:
\begin{equation}
\mathcal{O}_{\text{Haldane}} = \lim_{r \to \infty} \left\langle S^z_i \exp\left[i\pi \sum_{j=i+1}^{i+r-1} S^z_j \right] S^z_{i+r}  \right\rangle,
\end{equation}
with $r$ being the size of the string. Although ideally $r$ has to be large in our finite system, it will be a finite number.
In the region where \(\phi \in \left[\frac{1}{4}\pi, \frac{1}{2}\pi\right]\), the ground state is in the critical phase, which is a gapless phase. The existence of this phase can be reflected by a nematic structure factor at $q=2\pi / 3$,  which is defined as~\cite{Chiara2011} 
\[
S(q) = \frac{1}{N_d} \sum_{k, l} e^{iq(k-l)} \langle {(S^z_k)}^2 {(S^z_l)}^2 \rangle.
\]

The first step toward the VQE simulation of the Hamiltonian (\ref{eq:BBH_spin1}) is to generate its qubit representation. Interestingly, for spin-1 systems both binary and symmetry encodings demand exactly two qubits for representing each spin. This provides an opportunity to compare the performance of these two encodings using the same number of qubits. In order to map the Hamiltonian into a qubit basis, one has to use the transformation
$H=\mathcal{T}^{\otimes N_d} \widetilde{H} \mathcal{T}^{\dagger \otimes N_d}$, where $\mathcal{T}$ stands for $\mathcal{T}_b$ (for binary encoding) and $\mathcal{T}_s$ for symmetric encoding. Note that one has to pay extra attention to this transformation, in particular, when illegitimate states are also involved. This is because $\mathcal{T}\mathcal{T}^\dagger \neq \mathbb{I}_{\text{qubit}}$. Consequently,  
the transformation has to apply to the interaction terms \(\vec{\widetilde{S}}_i \cdot \vec{\widetilde{S}}_{i+1}\) and \((\vec{\widetilde{S}}_i \cdot \vec{\widetilde{S}}_{i+1})^2\), rather than directly on the spin-\(1\) operators \(\widetilde{S}^\alpha\) (with $\alpha=x,y,z$) and then use tensor product, where \(j = 2i - 1\) represents the corresponding index of the spin-\(1\) operator in the qubit system. The exact forms of the transformations for the symmetry and binary encodings are given in Appendixes~\ref{app:symmetry_encoding} and \ref{app:binary_encoding}, respectively.

It is insightful to check whether the ground state of the qubit Hamiltonian $\ket{E_1}$ truly represents the true qudit ground state $\ket{\widetilde{E}_1}$ through the fidelity $|\bra{E_1}\mathcal{T}\ket{\widetilde{E}_1}|^2$. In Fig.~\ref{fig:phase_diagram}(b), we present the fidelity result $|\bra{E_1}\mathcal{T}\ket{\widetilde{E}_1}|^2$ for the spin-$1$ BBH model with a system size of $N_d{=}6$, which corresponds to a $12$-qubit encoded Hamiltonian, across the phase diagram. 
Apparently, in the critical phase and part of the Haldane phase, this fidelity drops to zero showing that illegitimate states are indeed taking the low-energy states. Note that the result is valid for both binary and symmetry encoding methods. Therefore, without including penalty terms, see Eq.~(\ref{eq:penalized_cost_function}), one cannot target the true ground state.

\begin{figure*}[ht]
  \centering

  \begin{subfigure}[b]{0.49\columnwidth}
    \includegraphics[width=\linewidth]{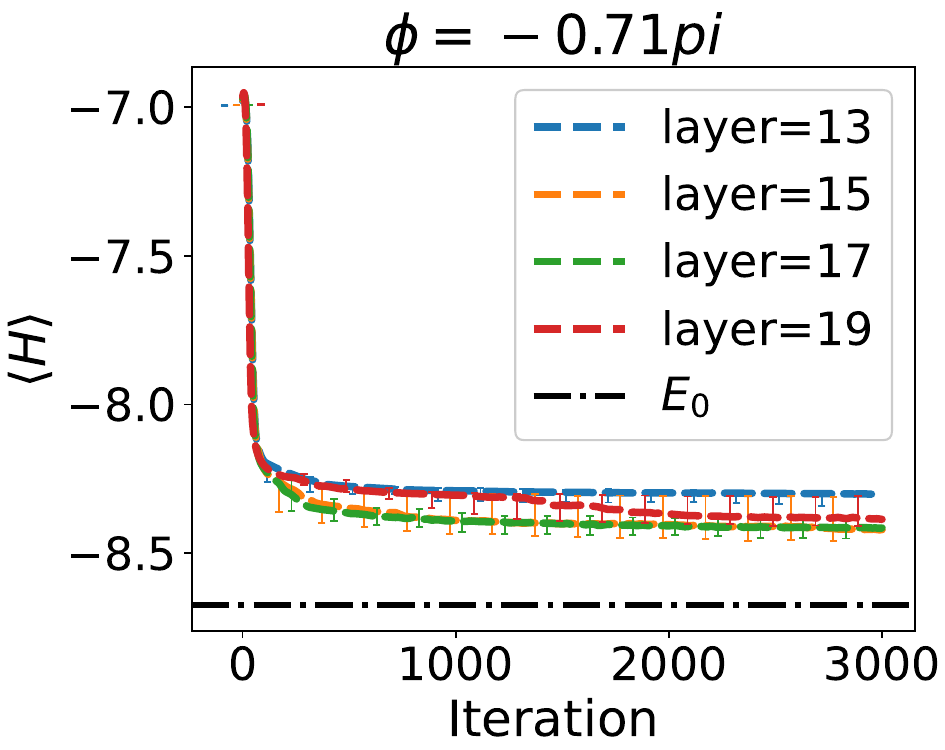}
    \caption{}
    \label{fig:binary_a}
  \end{subfigure}
  \hfill
  \begin{subfigure}[b]{0.49\columnwidth}
    \includegraphics[width=\linewidth]{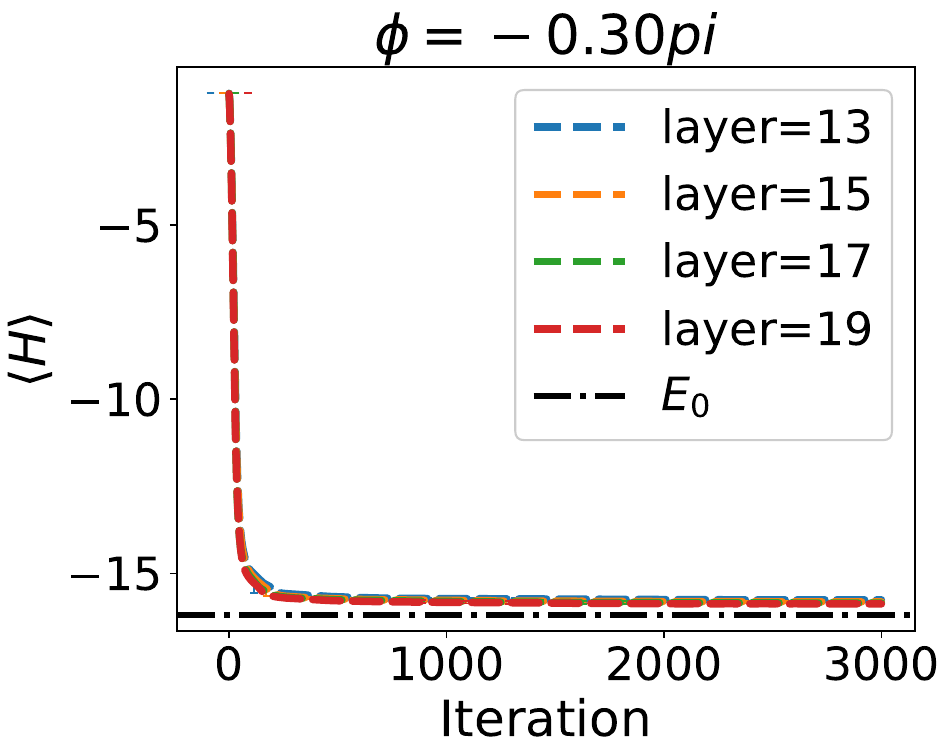}
    \caption{}    
    \label{fig:binary_b}
  \end{subfigure} 
  \begin{subfigure}[b]{0.49\columnwidth}
    \includegraphics[width=\linewidth]{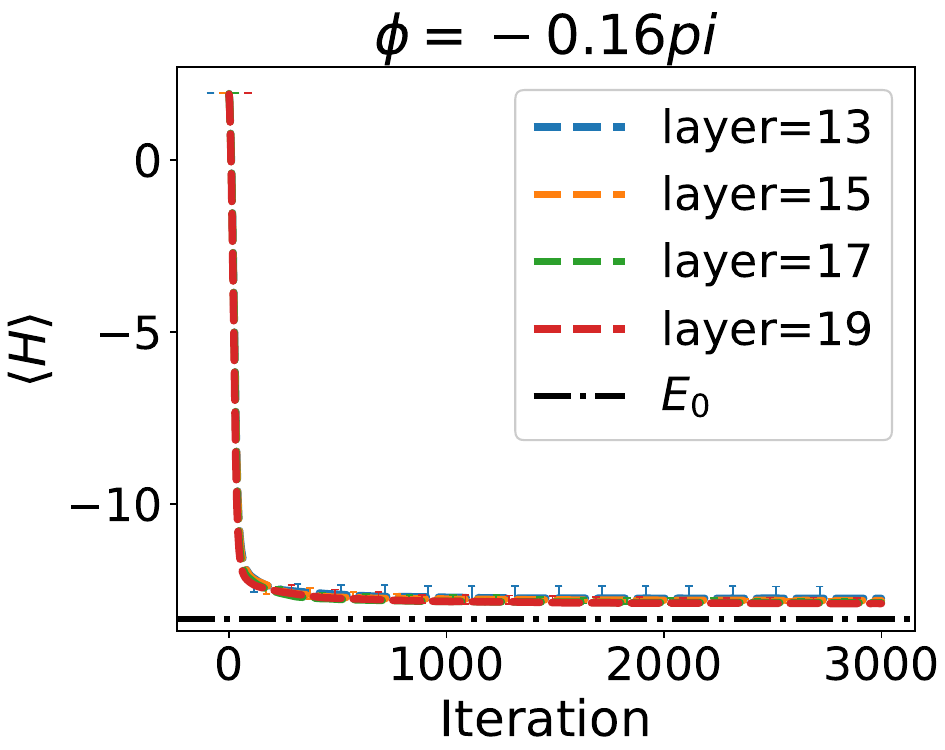}
    \caption{}    
    \label{fig:binary_c}
  \end{subfigure}
  \hfill
  \begin{subfigure}[b]{0.49\columnwidth}
    \includegraphics[width=\linewidth]{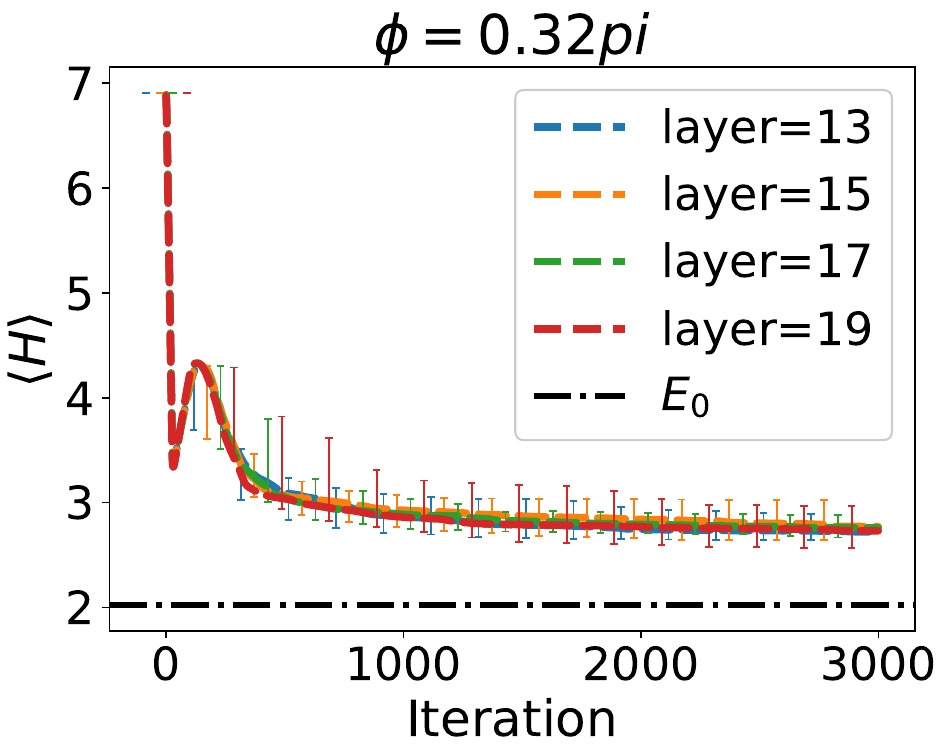}
    \caption{}
    \label{fig:binary_d} 
  \end{subfigure}\\

  \begin{subfigure}[b]{0.49\columnwidth}
    \includegraphics[width=\linewidth]{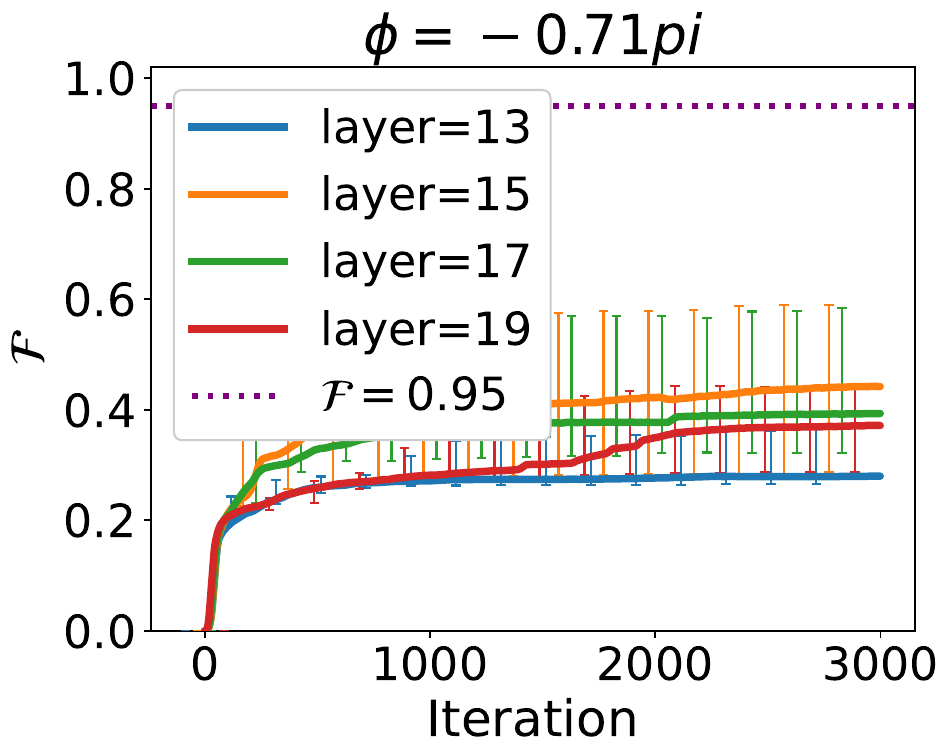}
    \caption{}
    \label{fig:binary_e}
  \end{subfigure}
  \hfill
  \begin{subfigure}[b]{0.49\columnwidth}
    \includegraphics[width=\linewidth]{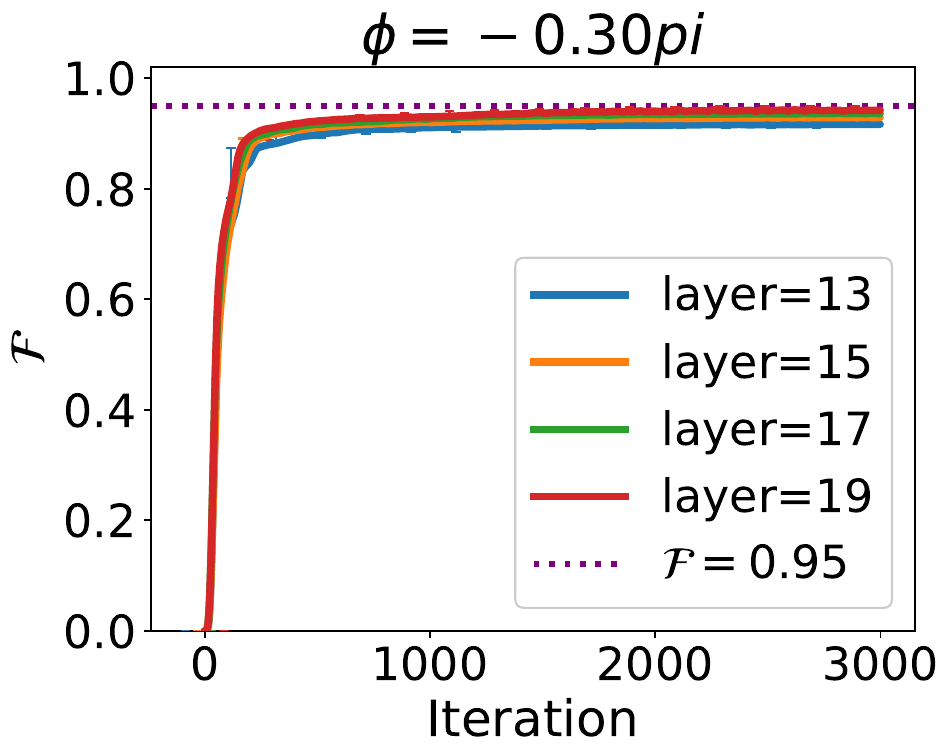}
    \caption{}    
    \label{fig:binary_f}
  \end{subfigure} 
  \begin{subfigure}[b]{0.49\columnwidth}
    \includegraphics[width=\linewidth]{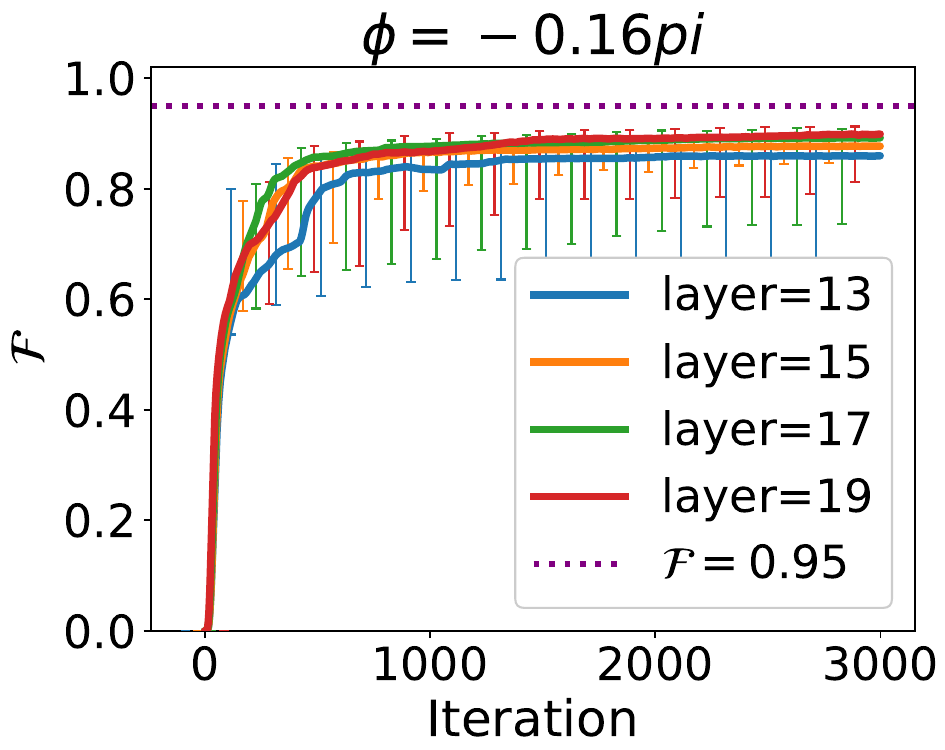}
    \caption{}    
    \label{fig:binary_g}
  \end{subfigure}
  \hfill
  \begin{subfigure}[b]{0.49\columnwidth}
    \includegraphics[width=\linewidth]{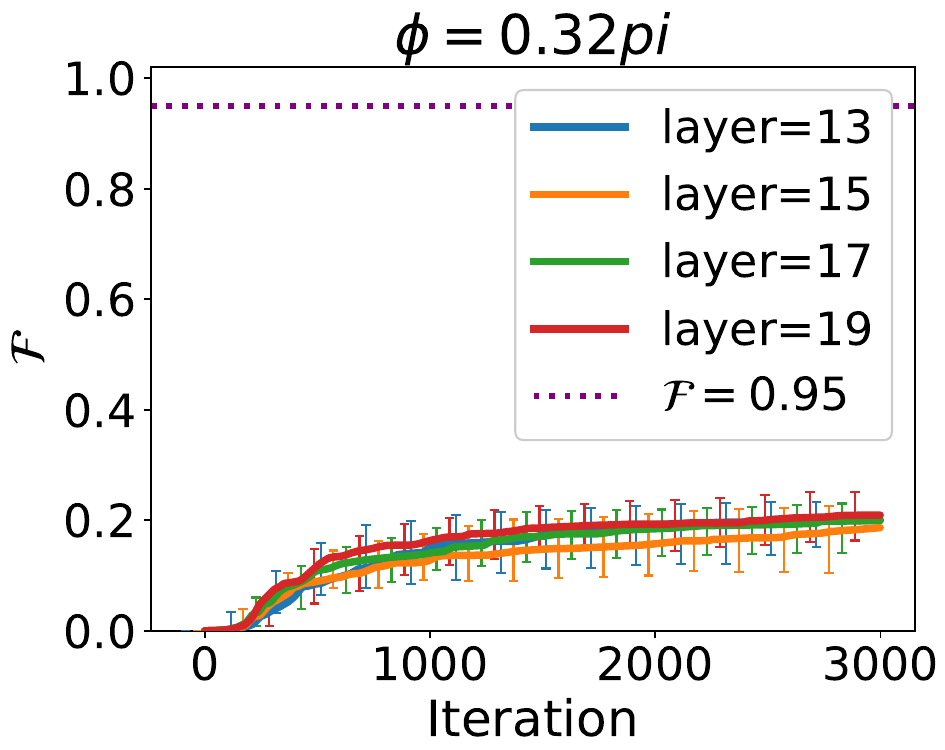}
    \caption{}
    \label{fig:binary_h} 
  \end{subfigure}\\

  \caption{
Binary encoding for the ground state of the spin-1 BBH model. The VQE simulation is shown for system size $N_d = 6$ and four representative values of $\phi$. The corresponding qubit Hamiltonian acts on 12 qubits encoded by the binary mapping. Each column corresponds to a fixed $\phi$: (a), (e) $\phi = -0.71\pi$ (dimerized); (b), (f) $\phi = -0.30\pi$ (dimerized); (c), (g) $\phi = -0.16\pi$ (Haldane); and (d), (h) $\phi = 0.32\pi$ (critical). The upper panels (a)--(d) show the energy expectation value $\langle H \rangle$ as a function of optimization iteration for different circuit depths, and the lower panels (e)--(h) show the corresponding median fidelity $\mathcal{F}$. The horizontal dash-dotted black lines indicate the exact ground-state energy $E_0$, and the horizontal purple dotted lines in panels (e)--(h) mark the target fidelity $\mathcal{F} = 0.95$. All curves are summarized over 50 independent runs with random initial parameters, and the error bars indicate the interquartile range (25th--75th percentile) across runs.
}

  \label{fig:binary}
\end{figure*}

In order to compare the binary and symmetry encodings, we first focus on VQE simulations of the ground state.
In both encodings, simulating a spin-1 chain of size $N_d$ requires $2N_d$ qubits. 
We start from the binary encoding, where each local spin is mapped to two qubits and a hardware-efficient circuit is used as the variational ansatz. 
The results for the BBH model with system size $N_d{=}6$ (a 12-qubit problem after encoding) are summarized in Fig.~\ref{fig:binary}. For four representative values of $\phi$ spanning the dimerized, Haldane, and critical regimes, we plot the energy expectation value and the corresponding fidelity with respect to the exact ground state as functions of the optimization iteration for several circuit depths. For each pair $(\phi,\text{layer})$, we perform 50 independent runs of the gradient-based Adam optimizer starting from random initial parameters; the curves show the median over runs, and the error bars indicate the interquartile range (IQR) (25th--75th percentiles).

Across all $\phi$ the energy converges rapidly to a plateau, while the obtainable fidelity depends sensitively on both $\phi$ and circuit depth. 
In the middle of the dimerized and Haldane phases ($\phi=-0.30\pi$ and $\phi=-0.16\pi$), moderate-depth circuits can reach fidelities close to the target threshold $\mathcal{F}=0.95$. 
In contrast, at the strongly dimerized point $\phi=-0.71\pi$ and near the critical point $\phi\approx0.32\pi$, the fidelity saturates well below this value even for the deepest ansatz considered, and increasing the depth beyond about 15 layers yields no improvement or even worse performance.
As quantified by the gradient norm analysis in Sec.~\ref{sec:barren_plateau}, this behavior is consistent with barren plateau trainability degradation in the binary-encoded hardware-efficient ansatz.


The VQE simulations with symmetry encoding utilize the total spin $s$ preserving ansatz, as illustrated in Fig.~\ref{fig:ansatzes}(b). In this case, it is crucial to select an initial state that aligns with the symmetry-preserving nature of the ansatz. Specifically, the ground state of the BBH model exhibits a total spin of $s{=}0$ across the dimerized, Haldane, and critical phases. Since the qubit Hamiltonian consists of $2N_d$ qubits, the initial state is taken to be $\ket{\psi_0} {=} \otimes^{N_d}\ket{\psi^-}$, where $\ket{\psi^-} = (\ket{01} - \ket{10})/\sqrt{2}$. This initial state naturally has a total spin of $s{=}0$. As we will see, unlike the binary encoding, the symmetry encoding method can faithfully simulate the ground state in all phases. It is worth emphasizing that the $\ket{\psi^-}$ is an illegitimate state for the qubits; however, as $\ket{\psi_0} {=} \otimes^{N_d}\ket{\psi^-}$ conserves the total spin, it eventually reaches the true ground state.

\begin{figure}[t]
  \centering

  \begin{subfigure}[b]{0.49\columnwidth}
    \includegraphics[width=\linewidth]{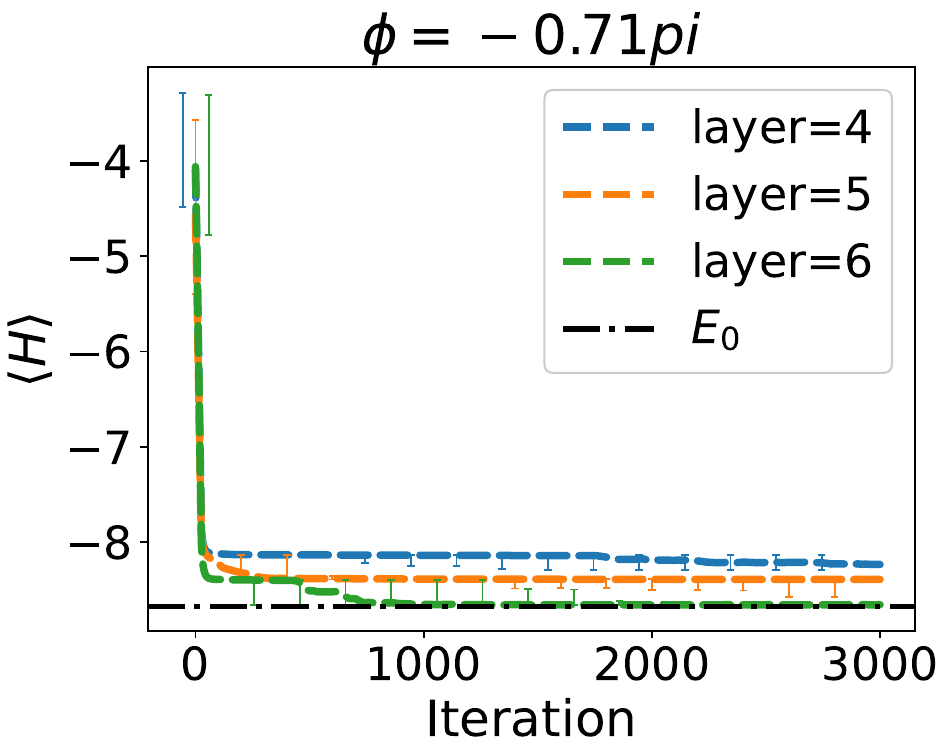}
    \caption{}
    \label{fig:dimer_a}
  \end{subfigure}
  \hfill
  \begin{subfigure}[b]{0.49\columnwidth}
    \includegraphics[width=\linewidth]{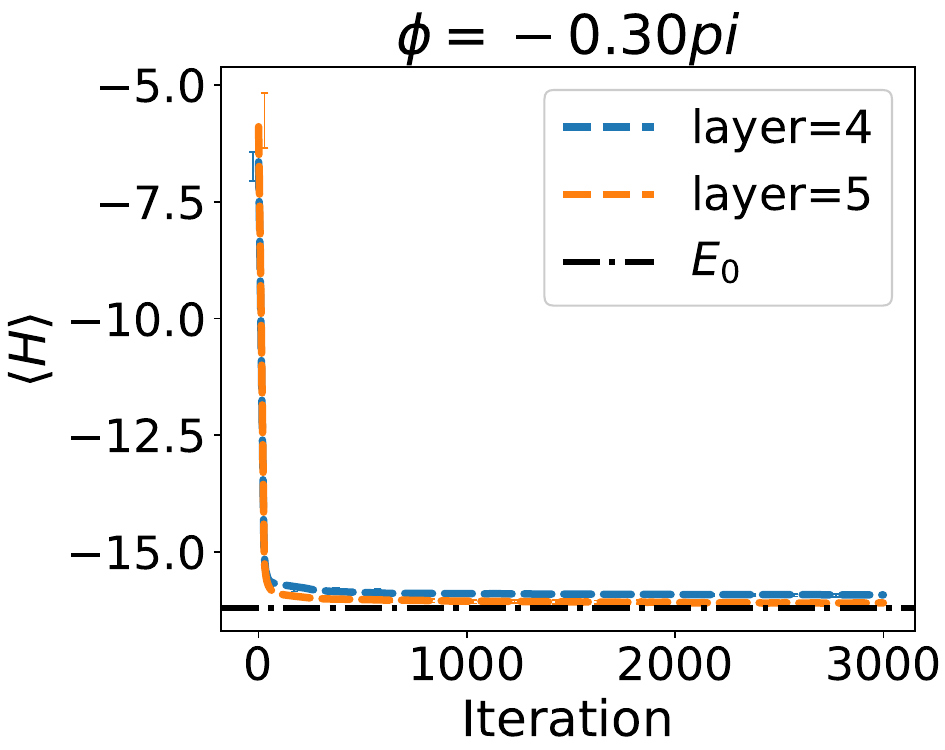}
    \caption{}    
    \label{fig:dimer_b}
  \end{subfigure} \\
  \begin{subfigure}[b]{0.49\columnwidth}
    \includegraphics[width=\linewidth]{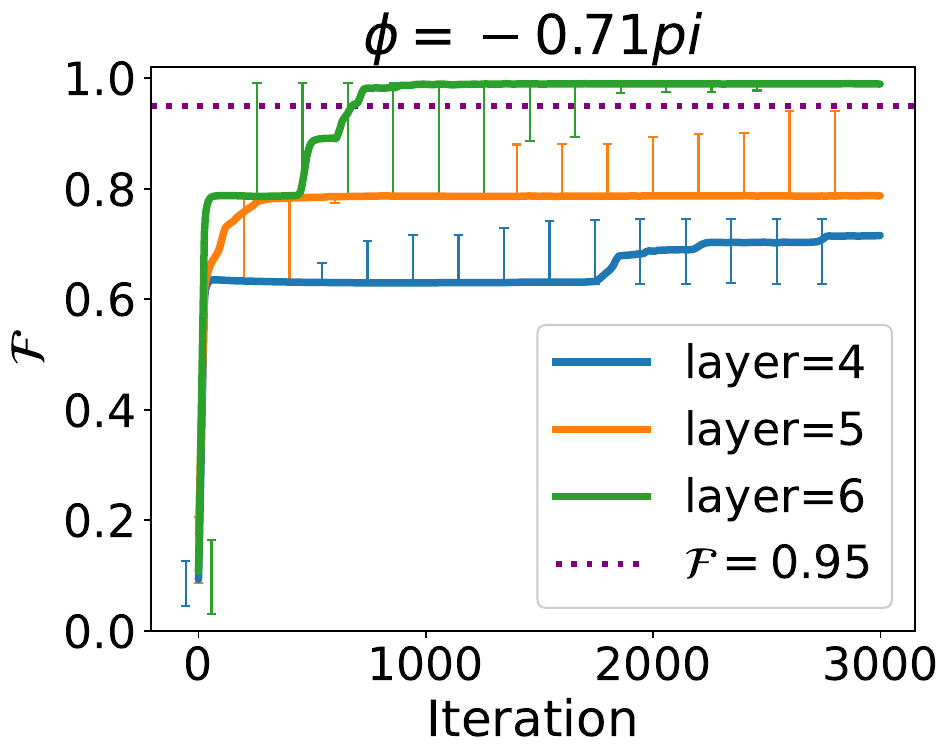}
    \caption{}
    \label{fig:dimer_c}
  \end{subfigure}
  \hfill
  \begin{subfigure}[b]{0.49\columnwidth}
    \includegraphics[width=\linewidth]{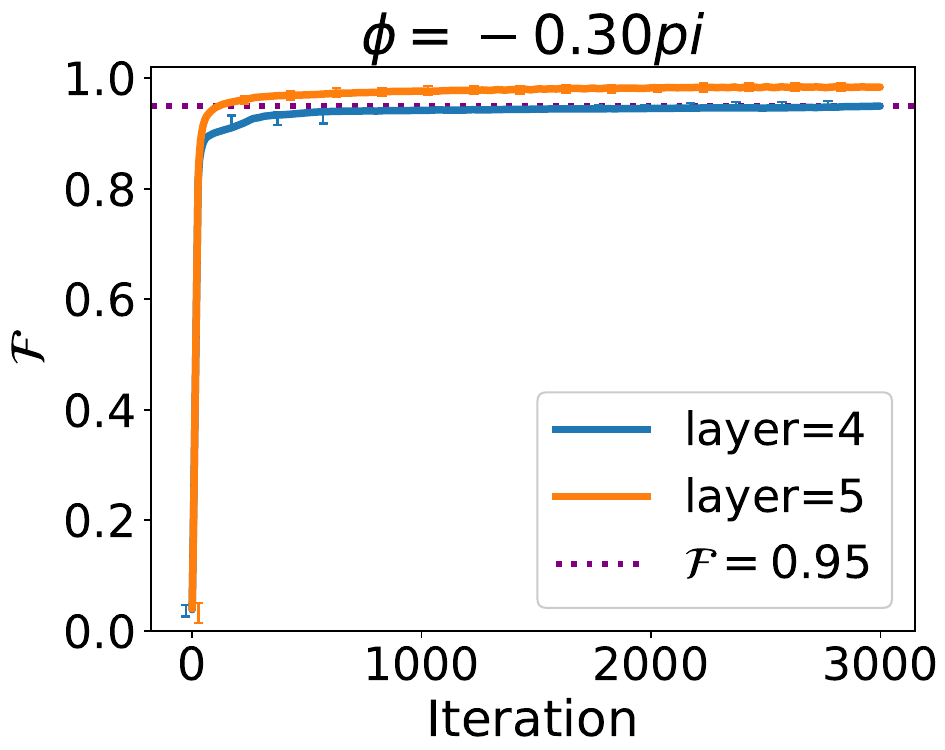}
    \caption{}
    \label{fig:dimer_d}
  \end{subfigure}

  \caption{
  Symmetry encoding in the dimerized phase of the spin-1 BBH model. The VQE simulation is shown for system size $N_d = 6$ and two values of $\phi$ in the dimerized phase. The corresponding qubit Hamiltonian acts on 12 qubits encoded by the symmetry-preserving mapping. The results are summarized over 50 independent runs with random initial parameters, and the error bars indicate the interquartile range (25th--75th percentile) across runs. The upper panels (a) and (b) show the energy expectation value $\langle H \rangle$ as a function of optimization iteration for $\phi = -0.71\pi$ and $\phi = -0.30\pi$, respectively. The lower panels (c) and (d) show the corresponding median fidelity $\mathcal{F}$.
}

  \label{fig:dimer_result}
\end{figure}

Let us consider the BBH model with system size $N_d=6$ at $\phi = -0.71\pi$ and $\phi = -0.30\pi$, both lying in the dimerized phase. 
In this regime, the ground state of the qubit Hamiltonian coincides with the true ground state of the original spin-1 model, so states outside the physical subspace do not affect the outcome and the cost function can simply be taken as the Hamiltonian expectation value $\langle H \rangle$.
The corresponding VQE results using the symmetry encoded, total spin preserving ansatz are shown in Figs.~\ref{fig:dimer_result}(a)--\ref{fig:dimer_result}(d), where the energy expectation value and fidelity are plotted as functions of the optimization iteration for different circuit depths.
For each pair $(\phi,\text{layer})$, we run the Adam optimizer from 50 random initializations and report the median over runs with interquartile-range error bars (25th--75th percentiles).
In both parameter regimes, the energy relaxes rapidly toward the exact ground-state value, and the obtainable fidelity exceeds $\mathcal{F} \approx 0.95$ after roughly $1.4\times10^3$ iterations using a six-layer ansatz for $\phi = -0.71\pi$ and a five-layer ansatz for $\phi = -0.30\pi$, demonstrating that the symmetry-encoded circuit can accurately prepare the ground state in the dimerized phase with substantially fewer layers than in the binary-encoded case.

\begin{figure}[t]
  \centering

  \begin{subfigure}[b]{0.49\columnwidth}
    \includegraphics[width=\linewidth]{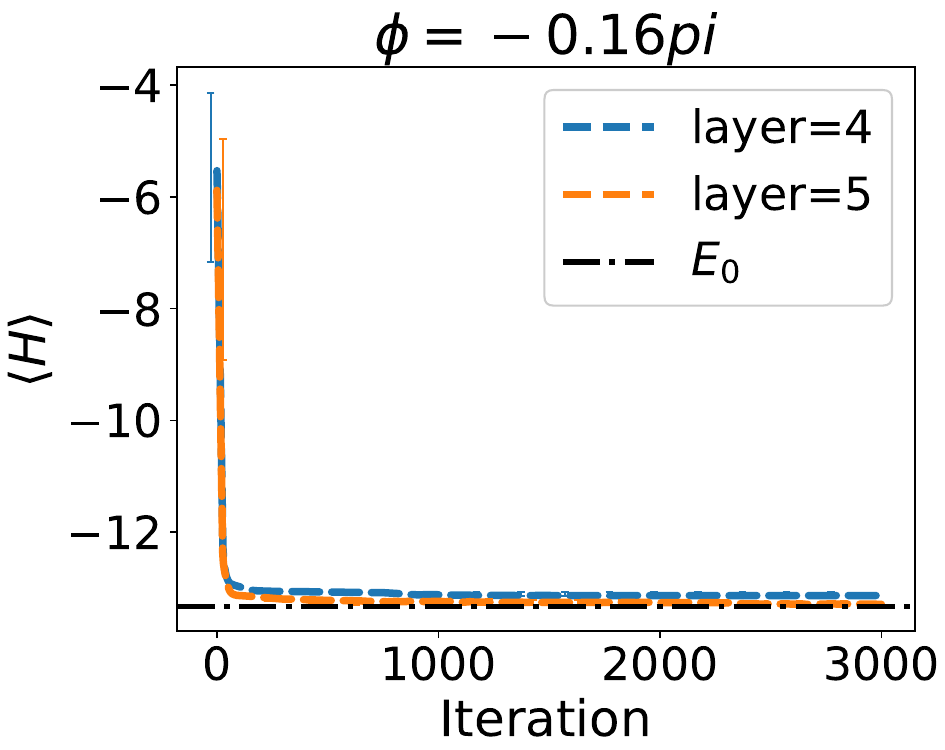}
    \caption{}
    \label{fig:haldane_a}
  \end{subfigure}
  \hfill
  \begin{subfigure}[b]{0.49\columnwidth}
    \includegraphics[width=\linewidth]{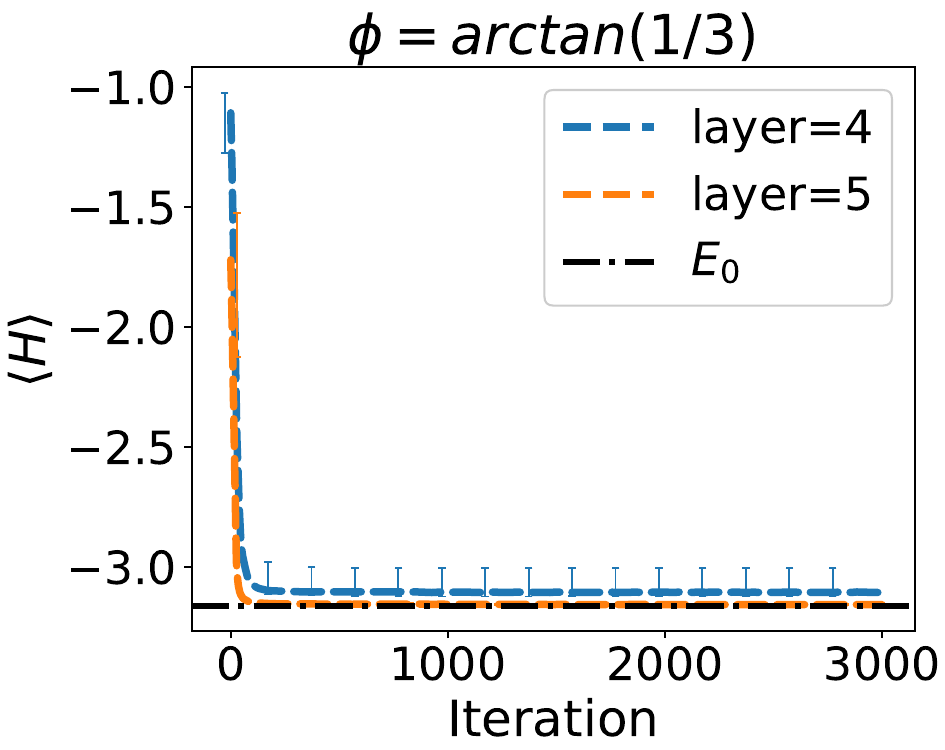}
    \caption{}
    \label{fig:haldane_b}
  \end{subfigure} \\
  \begin{subfigure}[b]{0.49\columnwidth}
    \includegraphics[width=\linewidth]{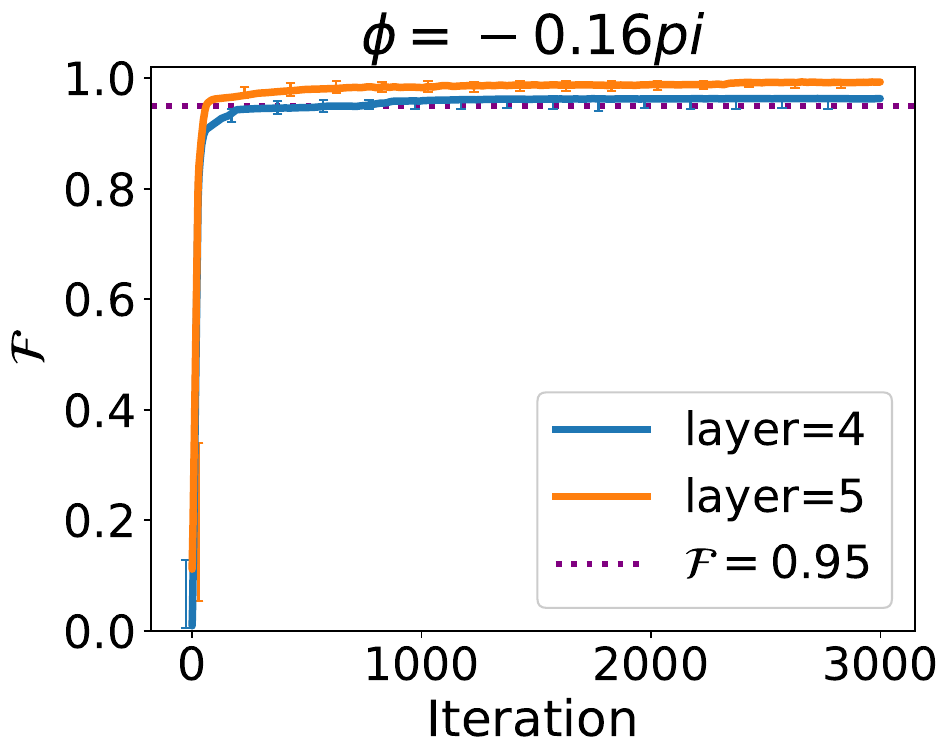}
    \caption{}
    \label{fig:haldane_c}
  \end{subfigure}
  \hfill
  \begin{subfigure}[b]{0.49\columnwidth}
    \includegraphics[width=\linewidth]{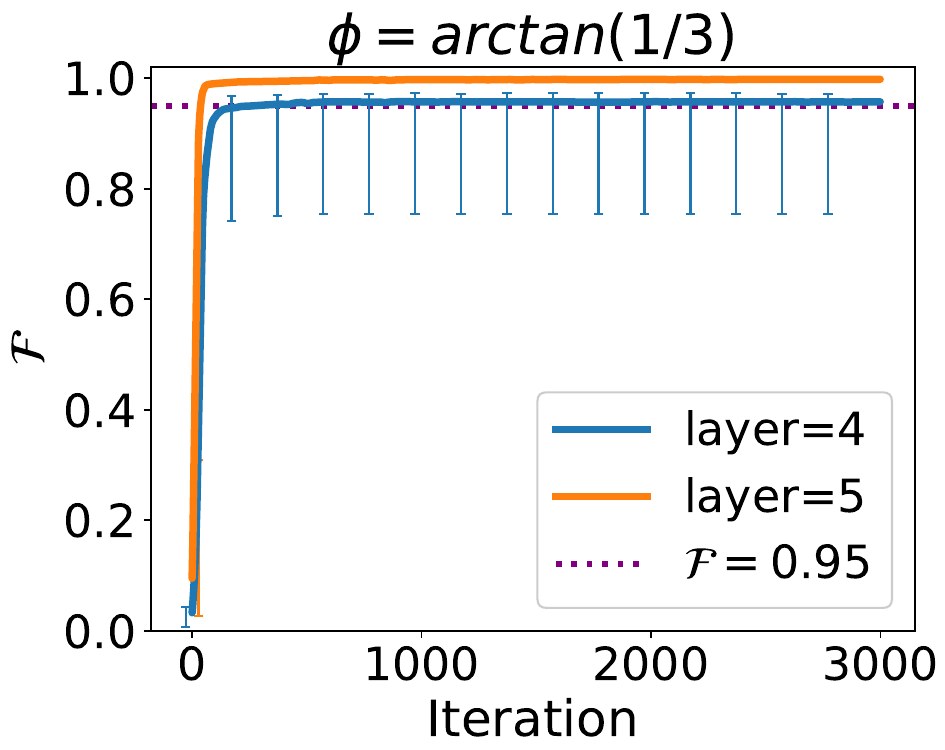}
    \caption{}
    \label{fig:haldane_d}
  \end{subfigure}

  \caption{
  Symmetry encoding in the Haldane phase of the spin-1 BBH model. The VQE simulation is shown for system size $N_d = 6$ and two values of $\phi$ in the Haldane phase. The corresponding qubit Hamiltonian acts on 12 qubits encoded by the symmetry-preserving mapping. The results are summarized over 50 independent runs with random initial parameters, and the error bars indicate the interquartile range (25th--75th percentile) across runs. The upper panels (a) and (b) show the energy expectation value $\langle H \rangle$ as a function of optimization iteration for $\phi = -0.16\pi$ and $\phi = \arctan(1/3)$ (the AKLT point), respectively. The lower panels (c) and (d) show the corresponding median fidelity $\mathcal{F}$.
}

  \label{fig:haldane_res}
\end{figure}

We now turn to the Haldane phase of the BBH model, focusing on $\phi = -0.16\pi$ and $\phi = \arctan(1/3)$. 
As in the dimerized phase, in this regime, the ground state of the qubit Hamiltonian coincides with the true ground state of the original spin-1 model, so the illegitimate states are pushed to high energies and do not affect the low-energy physics.
Consequently, the cost function can again be taken as the Hamiltonian expectation value $\langle H \rangle$.
At $\phi = \arctan(1/3)$, the AKLT point, the ground space is fourfold degenerate, containing a singlet with total spin $s=0$ and a triplet with $s=1$.
Using a total spin preserving ansatz, one can target any of these states by choosing an appropriate initial state with fixed $s$. 
For example, the $s{=}0$ ground state can be reached from the product of singlets $\ket{\psi_0} = \bigotimes^{N_d}\ket{\psi^-}$, while the $s{=}1$ ground states can be obtained by replacing one singlet with $\ket{\psi^+}=(\ket{01}+\ket{10})/\sqrt{2}$ (for $s_z=0$) or $\ket{\phi^{\pm}}=(\ket{00}\pm\ket{11})/\sqrt{2}$ (for $s_z=\pm 1$).
The VQE results with the symmetry encoded, spin preserving ansatz are shown in Figs.~\ref{fig:haldane_res}(a)--\ref{fig:haldane_res}(d), where the energy expectation value and fidelity are plotted as functions of the optimization iteration for different circuit depths.
For each $(\phi,\text{layer})$, we perform 50 independent runs of the Adam optimizer initialized with random parameters, and we report the median over runs with interquartile-range error bars (25th--75th percentiles).
In both cases, a five-layer ansatz is sufficient to reach fidelities $\mathcal{F} \gtrsim 0.95$, with convergence occurring after approximately $1.75\times 10^3$ iterations for $\phi = -0.16\pi$ and $1.4\times 10^3$ iterations for $\phi = \arctan(1/3)$.
These results demonstrate that the symmetry encoded VQE can accurately capture the Haldane phase, including the highly entangled AKLT point, with relatively shallow circuits.

\begin{figure}[t]
  \centering
  \begin{minipage}[b]{0.49\columnwidth}
    \centering
    \includegraphics[width=\linewidth]{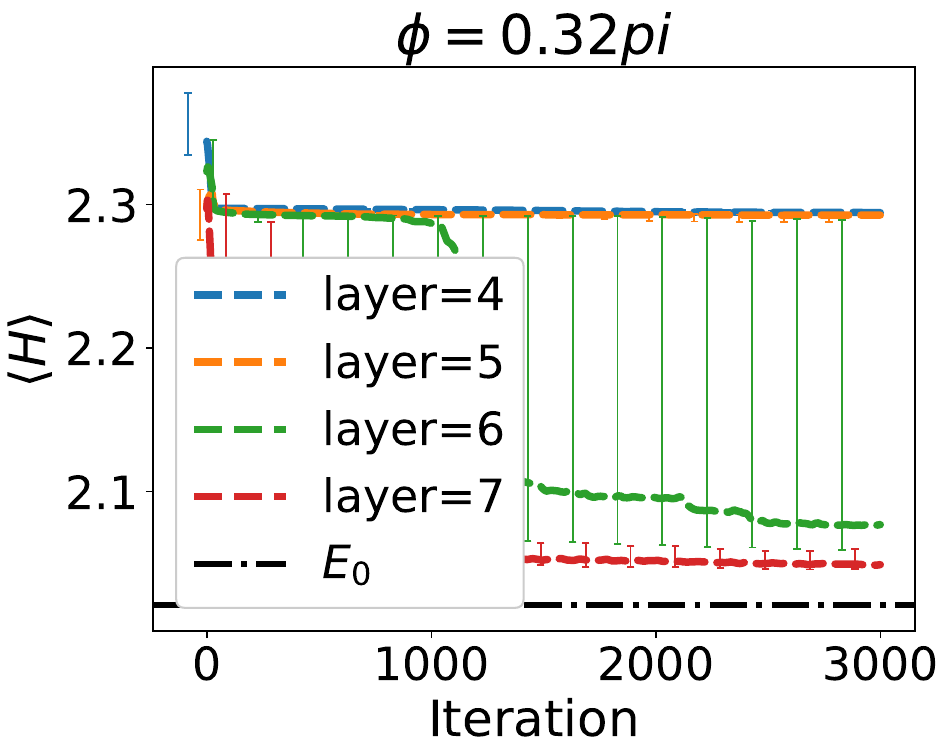}
    \par\smallskip (a)
  \end{minipage}
  \hfill
  \begin{minipage}[b]{0.49\columnwidth}
    \centering
    \includegraphics[width=\linewidth]{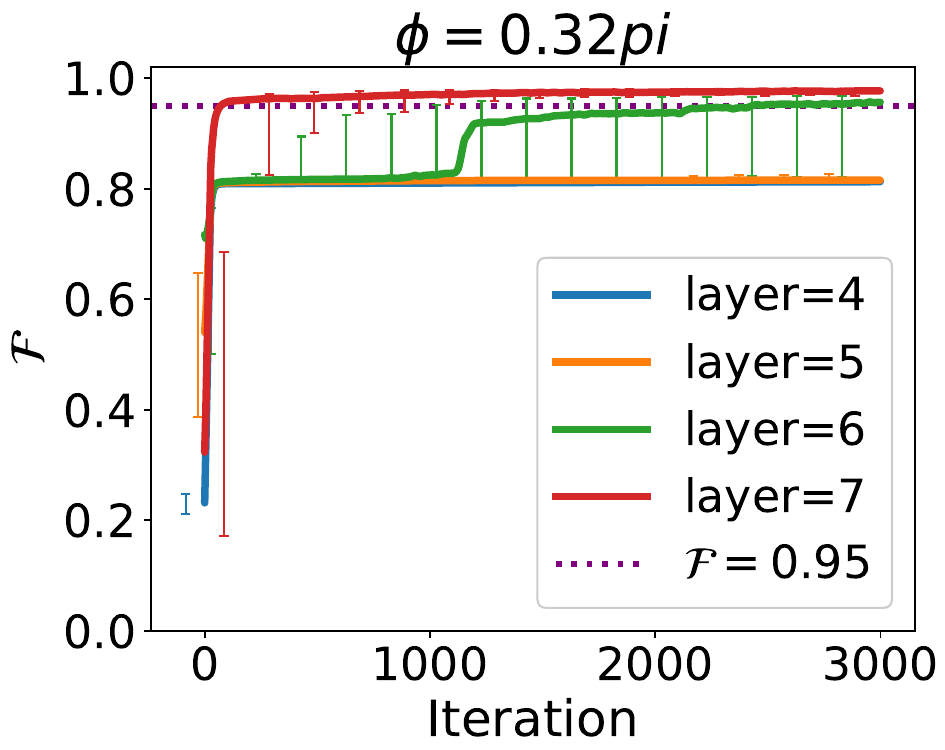}
    \par\smallskip (b)
  \end{minipage}
  \caption{
  Symmetry encoding in the critical phase of the spin-1 BBH model. The VQE simulation is shown for system size $N_d = 6$ at $\phi = 0.32\pi$. The corresponding qubit Hamiltonian acts on 12 qubits encoded by the symmetry-preserving mapping. The results are summarized over 50 independent runs with random initial parameters, and the error bars indicate the interquartile range (25th--75th percentile) across runs. Panel (a) shows the energy expectation value $\langle H \rangle$ as a function of optimization iteration, and panel (b) shows the corresponding median fidelity $\mathcal{F}$.
}

  \label{fig:critical_res}
\end{figure}

We further explore the critical phase of the BBH model at $\phi = 0.32\pi$, a regime in which illegitimate states strongly affect the low-energy spectrum. For $N_d=6$, the true ground state of the original spin-1 Hamiltonian corresponds to the 1228th eigenstate of the qubit Hamiltonian when its eigenvalues are ordered in ascending energy. To steer the optimization toward the physical ground state, we augment the VQE cost with the penalty term defined in Eqs.~\eqref{eq:spin1_penalty_term} and \eqref{eq:penalized_cost_function}, using $\beta = 1$. The resulting VQE performance for different circuit depths is shown in Figs.~\ref{fig:critical_res}(a) and \ref{fig:critical_res}(b), where we plot the energy expectation value and fidelity versus optimization iteration. For each depth, we run the Adam optimizer from 50 random initial parameters and report the median with interquartile-range error bars (25th--75th percentiles). These simulations show that a six-layer symmetry-preserving ansatz is sufficient to reach fidelities $\mathcal{F} \gtrsim 0.95$, reliably preparing the true ground state despite its high index in the qubit spectrum. As expected, the presence of the penalty term increases the optimization cost, leading to a larger number of iterations compared with the dimerized and Haldane phases; nevertheless, the penalized VQE successfully captures the critical ground state of the BBH model.

A detailed comparison of the quantum and classical resources, as well as the performance metrics (fidelity and iterations), between the binary and symmetry encoding methods for the spin-1 model is provided in Table~\ref{tab:spin1_results}. The data confirm that the binary encoding fails to converge to the ground state in most phases, whereas the symmetry encoding reliably achieves high fidelity with substantially lower classical cost. The two-qubit-gate counts, however, show a more nuanced trade-off: Symmetry encoding uses fewer CNOT gates in three of the four phases listed in Table~\ref{tab:spin1_results}, but more CNOT gates at \(\phi=-0.71\pi\). Importantly, even in that case the smaller CNOT count of the binary encoding does not translate into a competitive result, since the achieved fidelity remains far below that of the symmetry-preserving ansatz. Thus, the main advantage of symmetry encoding in this benchmark is not a reduction in gate count, but rather the combination of improved fidelity and trainability made possible by the symmetry-preserving ansatz. This contrast is consistent with the gradient norm diagnostic in Sec.~\ref{sec:barren_plateau}, which shows much stronger gradient suppression for the binary hardware-efficient ansatz than for the symmetry-preserving ansatz.

\begin{figure*}[ht]
  \centering
  \begin{minipage}[b]{0.328\textwidth}
    \centering
    \includegraphics[width=\linewidth]{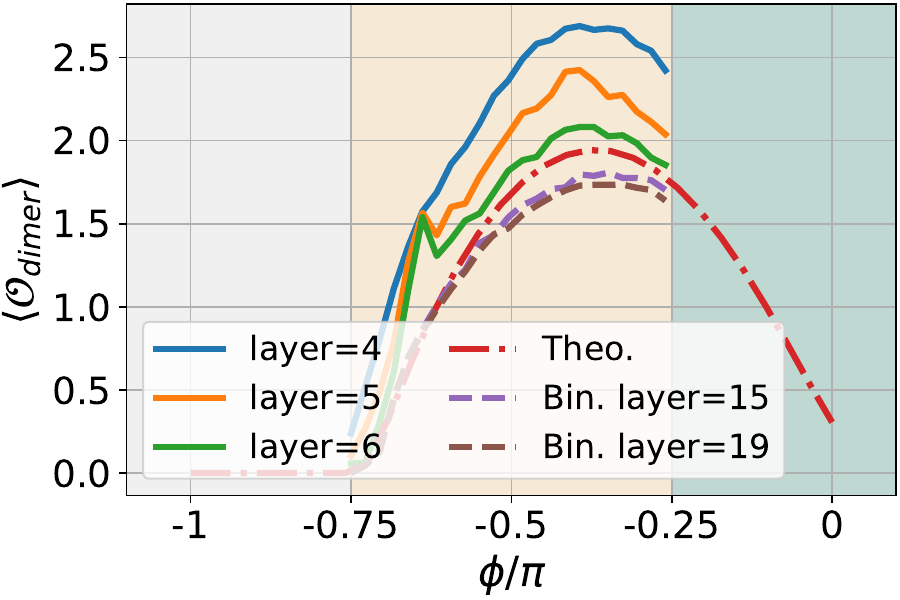}
    \par\smallskip (a)
  \end{minipage}
  \hfill
  \begin{minipage}[b]{0.328\textwidth}
    \centering
    \includegraphics[width=\linewidth]{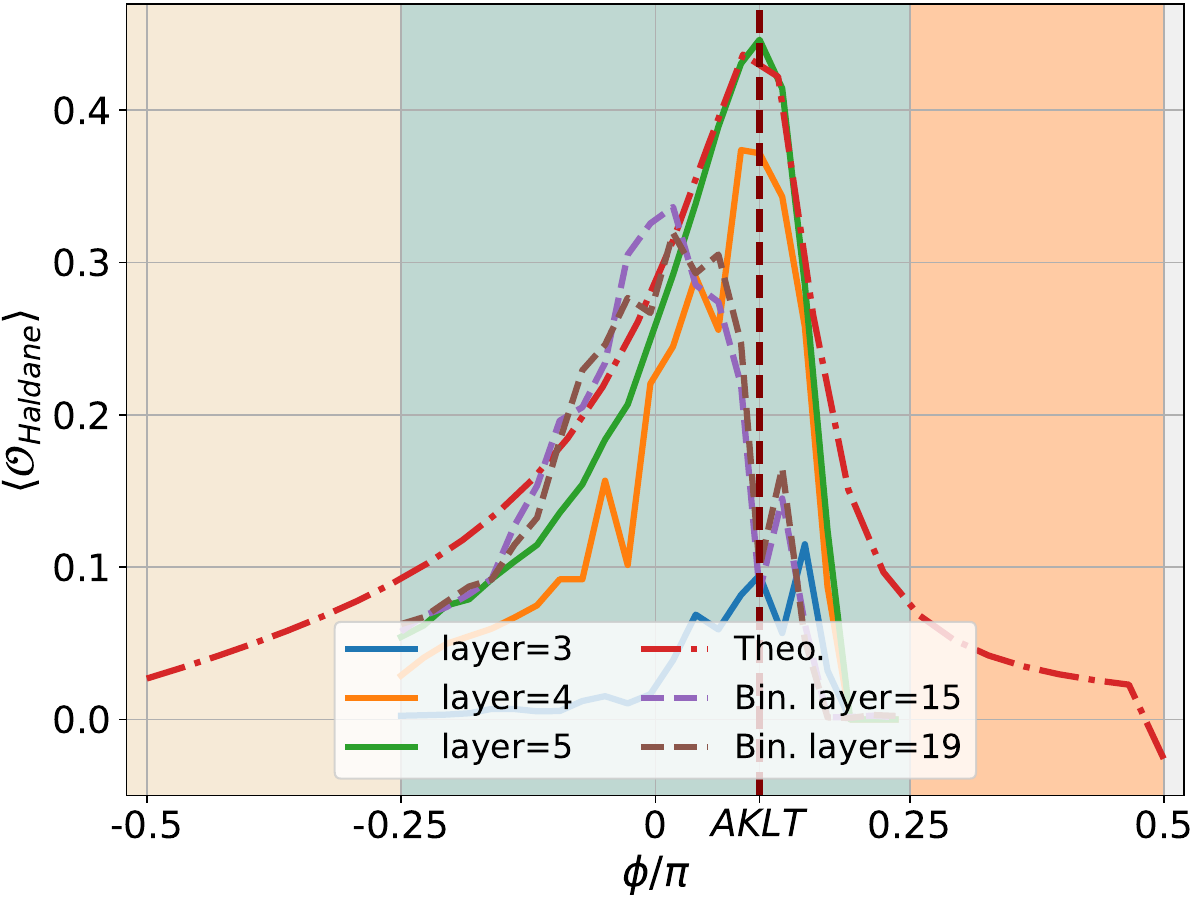}
    \par\smallskip (b)
  \end{minipage}
  \hfill
  \begin{minipage}[b]{0.328\textwidth}
    \centering
    \includegraphics[width=\linewidth]{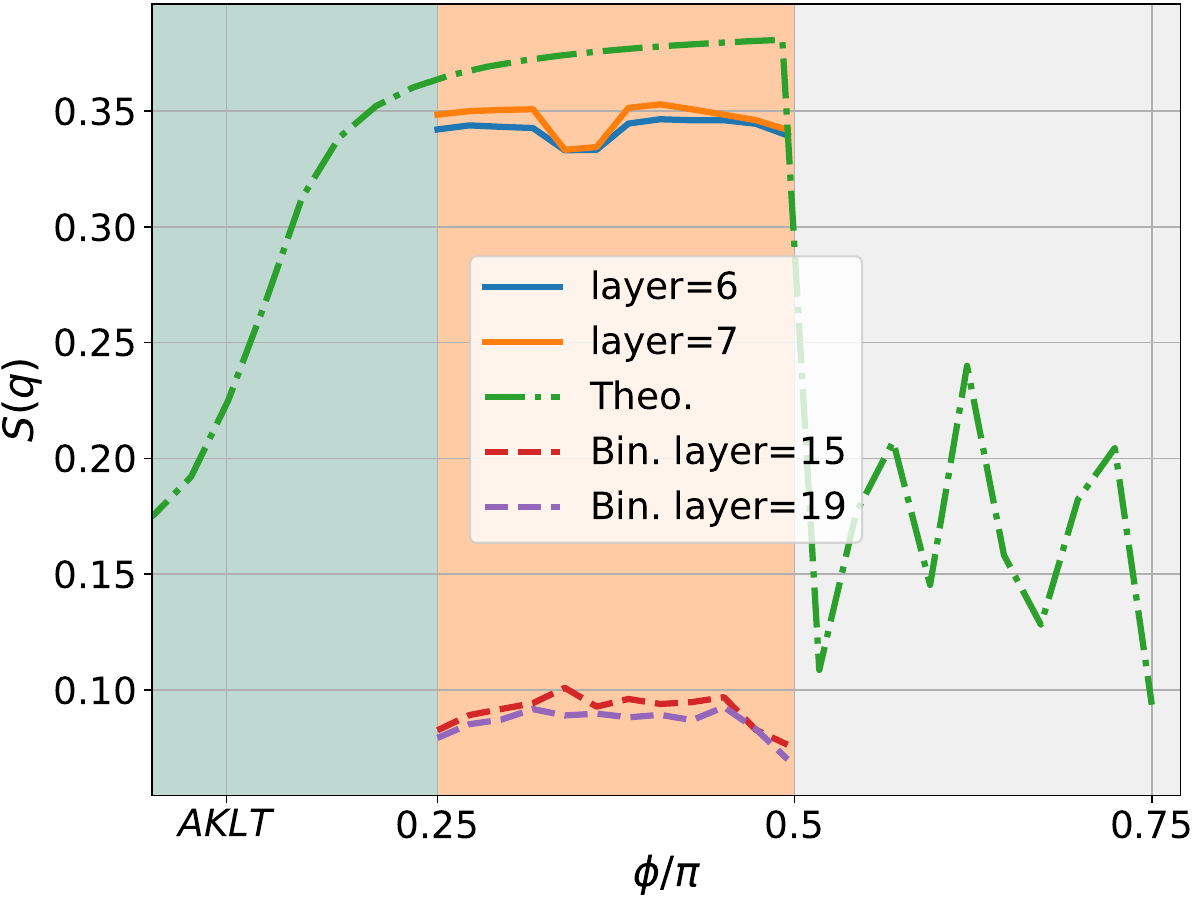}
    \par\smallskip (c)
  \end{minipage}
  \caption{Order parameters across the phases of the spin-1 BBH model. Panels (a)--(c) show the order parameters extracted from the VQE output states for the dimerized, Haldane, and critical phases, respectively. The results are summarized over 50 independent runs. The red dash-dotted curves denote the theoretical predictions. As the number of layers in the total-spin-preserving ansatz increases, the values obtained with symmetry encoding approach the theoretical predictions, whereas binary encoding fails to reproduce the correct phase behavior.}
  \label{fig:order_para_res}
\end{figure*}

As discussed above, each phase of the BBH model is characterized by a distinct order parameter. We evaluate these order parameters on the VQE-prepared ground states, and the results are shown in Figs.~\ref{fig:order_para_res}(a)--\ref{fig:order_para_res}(c). We observe that the symmetry encoding accurately captures the order parameters as the circuit depth increases, aligning with the theoretical predictions. In stark contrast, the binary encoding fails to converge to the ground state, and as shown in the figure, it fails to capture the correct order parameters. The successful reproduction of these order parameters by the symmetry encoding confirms its capability to capture the ground-state properties of the BBH model across the dimerized, Haldane, and critical phases.

The observed behavior across $\phi$ reproduces the expected signatures of the dimerized, Haldane, and critical phases, confirming that our VQE simulations correctly capture the ground-state properties of the BBH model throughout the phase diagram.

\begin{table}[ht]
  \footnotesize
  \centering
  \setlength{\tabcolsep}{3pt}
  \resizebox{\columnwidth}{!}{%
  \begin{tabular}{@{}lcccc@{}}
  \toprule
  \multicolumn{5}{c}{Binary encoding, \( N_q=12 \)} \\
  \cmidrule(r){1-5}
  & {\( \phi = -0.71\pi \)} & {\( \phi = -0.30\pi \)} & {\( \phi = -0.16\pi \)} & {\( \phi = 0.32\pi \)}\\
  No. of CNOT    & 187 & 198 & 187 & 264 \\
  No. of parameters   & 432 & 456 & 432 & 600 \\
  \( \mathcal{F} \) & 0.655 $\pm$ 0.056 & 0.944 $\pm$ 0.015 & 0.897 $\pm$ 0.042 & 0.313 $\pm$ 0.423\\
  Iteration & 3914 $\pm$ 2092 & 4740 $\pm$ 1742 & 6646 $\pm$ 3152 & 4974 $\pm$ 2315\\
  $C_R$ ($\times 10^3$) & 1690.6 $\pm$ 904.0 & 2161.2 $\pm$ 794.4 & 2871.3 $\pm$ 1361.7 & 2984.1 $\pm$ 1389.2\\
  \midrule
  \multicolumn{5}{c}{Symmetry encoding, \( N_q=12 \)} \\
  \cmidrule(r){1-5}
  & {\( \phi = -0.71\pi \)} & {\( \phi = -0.30\pi \)} & {\( \phi = -0.16\pi \)} & {\( \phi = 0.32\pi \)}\\
  No. of CNOT    & 204 & 138 & 138 & 171 \\
  No. of parameters   & 66 & 44 & 44 & 55 \\
  \( \mathcal{F} \) & 0.991 $\pm$ 0.001 & 0.955 $\pm$ 0.036 & 0.953 $\pm$ 0.027 & 0.953 $\pm$ 0.118\\
  Iteration & 666 $\pm$ 710 & 530 $\pm$ 279 & 605 $\pm$ 176 & 1462 $\pm$ 634\\
  $C_R$ ($\times 10^3$) & 43.9 $\pm$ 46.8 & 23.3 $\pm$ 12.3 & 26.6 $\pm$ 7.7 & 80.4 $\pm$ 34.9\\
  \bottomrule
  \end{tabular}
  }
  \caption{Comparison between binary and symmetry encoding methods for spin-1 model with \(N_d=6\) (12 qubits). The table shows the number of CNOT gates, the number of parameters, median fidelity (\(\mathcal{F}\)), median iteration count, and median classical resources ($C_R$), with interquartile range for fidelity, iteration and classical resource metrics. For binary encoding, the reported resources correspond to the layer with the highest median fidelity in each phase, since \(\mathcal{F} < 0.95\) for all tested depths. For symmetry encoding, the reported resources correspond to the shallowest layer reaching \(\mathcal{F} \geq 0.95\).}
  \label{tab:spin1_results}
\end{table}

\begin{table}[ht]
  \footnotesize
  \centering
  \setlength{\tabcolsep}{3pt}
  \resizebox{\columnwidth}{!}{%
  \begin{tabular}{@{}lccc@{}}
  \toprule
  \multicolumn{4}{c}{Binary encoding, \( N_q = 12 \)} \\
  \cmidrule(r){1-4}
  & {\( \phi = -0.75\pi \)} & {\( \phi = -0.36\pi \)} & {\( \phi = -0.15\pi \)} \\
  No. of CNOT    & 132 & 132 & 132 \\
  No. of parameters   & 312 & 312 & 312 \\
  \( \mathcal{F} \) & 0.432 $\pm$ 0.417 & 0.806 $\pm$ 0.208 & 0.513 $\pm$ 0.256 \\
  Iteration & 3076.4 $\pm$ 1082.7 & 2785.3 $\pm$ 1201.8 & 3244.2 $\pm$ 1211.9 \\
  $C_R$ ($\times 10^3$) & 959.8 $\pm$ 337.8 & 869.0 $\pm$ 375.0 & 1012.2 $\pm$ 378.1 \\
  \midrule
  \multicolumn{4}{c}{Symmetry encoding, \( N_q = 18 \)} \\
  \cmidrule(r){1-4}
  & {\( \phi = -0.75\pi \)} & {\( \phi = -0.36\pi \)} & {\( \phi = -0.15\pi \)} \\
  No. of CNOT    & 612 & 459 & 612 \\
  No. of parameters   & 204 & 153 & 204 \\
  \( \mathcal{F} \) & 0.970 $\pm$ 0.008 & 0.951 $\pm$ 0.010 & 0.973 $\pm$ 0.011 \\
  Iteration & 8806 $\pm$ 1839.7 & 4888.7 $\pm$ 1479.5 & 6958.2 $\pm$ 1712.9 \\
  $C_R$ ($\times 10^3$) & 1796.4 $\pm$ 375.3 & 748.0 $\pm$ 226.4 & 1419.5 $\pm$ 350.4 \\
  \bottomrule
  \end{tabular}
  }
  \caption{Comparison between binary and symmetry encoding methods for spin $s=3/2$ at three different values of \(\phi\) for \(N_q = 12\) and \(N_q = 18\). The table shows the number of CNOT gates, the number of parameters, median fidelity (\(\mathcal{F}\)), median iteration count, and median classical resources ($C_R$), with interquartile range for fidelity, iteration, and classical resource metrics.}
  \label{tab:spin3/2_results}
\end{table}



\section{Spin-3/2 bilinear-biquadratic Heisenberg model}

For spin-\(1\) systems, where both binary and symmetry encodings require the same number of qubits, the result for both cases shows a clear advantage on the symmetry encoding. This advantage is expected as exploiting symmetries in VQE simulations reduces the search subspace in the Hilbert subspace to only the relevant part in which all the states have the desired symmetries. This matches with the previous results for VQE simulation of qubit systems~\cite{Lyu2023symmetryenhanced}. 
By increasing $d$, however, the binary and symmetry encoding methods require different numbers of qubits. In this situation, it is not clear whether the symmetry encoding still performs better as it demands more qubits for the mapping. To explore this situation, we focus on a spin-$3/2$ model. We consider the same Hamiltonian as the spin-$1$ BBH model, shown in Eq.~\eqref{eq:BBH_spin1}, but with the spin-1 operators replaced by spin-$3/2$ operators. The corresponding spin-$3/2$ operators are represented by~matrices:
\begin{align*}
  \widetilde{S}^x &= \frac{1}{2} \begin{pmatrix}
    0 & \sqrt{3} & 0 & 0 \\
    \sqrt{3} & 0 & 2 & 0 \\
    0 & 2 & 0 & \sqrt{3} \\
    0 & 0 & \sqrt{3} & 0
  \end{pmatrix}, \\
  \widetilde{S}^y &= \frac{1}{2i} \begin{pmatrix}
    0 & \sqrt{3} & 0 & 0 \\
    -\sqrt{3} & 0 & 2 & 0 \\
    0 & -2 & 0 & \sqrt{3} \\
    0 & 0 & -\sqrt{3} & 0
  \end{pmatrix}, \\
  \widetilde{S}^z &= \frac{1}{2} \begin{pmatrix}
    3 & 0 & 0 & 0 \\
    0 & 1 & 0 & 0 \\
    0 & 0 & -1 & 0 \\
    0 & 0 & 0 & -3
  \end{pmatrix}.
\end{align*}
The spin-\(3/2\) BBH model is a four-level system with the basis determined by the eigenvalues of the $S_z$ operator. The basis is then given by
\begin{eqnarray}
    \ket{\widetilde{0}}&=&\ket{s_z{=}{+}\frac{3}{2}}, \quad 
    \ket{\widetilde{1}}=\ket{s_z{=}{+}\frac{1}{2}}, \cr
    \ket{\widetilde{2}}&=&\ket{s_z{=}{-}\frac{1}{2}}, \quad
    \ket{\widetilde{3}}=\ket{s_z{=}{-}\frac{3}{2}}.  
\end{eqnarray}

{ Spin-$3/2$ models provide a natural setting for studying four-level quantum systems. They arise in high-spin cold atomic systems and in SU($4$) spin-orbital physics. In optical lattices, fermions with hyperfine spin $F=3/2$ can exhibit enlarged SO($5$) or SU($4$) symmetries and competing quantum phases~\cite{Wu2006spin32,Yamashita1998}. Generalized spin-$3/2$ Heisenberg chains with bilinear, biquadratic, and higher-order interactions have also been used to study quadrupolar orders, dimerization, and tetramerization~\cite{Chen2023spin32}. Therefore, the spin-$3/2$ BBH model provides a physically motivated four-level test case for comparing the two encodings when they require different numbers of qubits.}

To encode these states into qubits using binary encoding, one requires two qubits as
\begin{align}\label{eq:spin3/2_mapping_binary}
  &\ket{\widetilde{0}} \mapsto \ket{00}, \quad \ket{\widetilde{1}} \mapsto \ket{01}, \\ \nonumber 
  & \ket{\widetilde{2}} \mapsto \ket{10}, \quad \ket{\widetilde{3}} \mapsto \ket{11}.
\end{align}
On the other hand, the symmetry encoding method of the spin-\(3/2\) BBH model requires three qubits for each spin. The mapping for the symmetry encoding method is as follows:
\begin{align}\label{eq:spin3/2_mapping_symm}
  &\ket{\widetilde{0}} \mapsto \ket{000}, \quad \ket{\widetilde{1}} \mapsto \frac{1}{\sqrt{3}}(\ket{001} + \ket{010} + \ket{100}), \\ \nonumber 
  & \ket{\widetilde{2}} \mapsto \frac{1}{\sqrt{3}}(\ket{110} + \ket{101} + \ket{011}), \quad \ket{\widetilde{3}} \mapsto \ket{111}.
\end{align}
Thus, exploiting the symmetry in the spin-\(3/2\) BBH model comes at the cost of using more qubits with the symmetry encoding method. Here, we compare the performance of binary encoding and symmetry encoding methods for VQE simulation of the spin-\(3/2\) BBH model. 
In Table~\ref{tab:spin3/2_results}, we present VQE simulation results for this model with $N_d=6$ spins, which is mapped to qubit systems of size $2N_d$ and $3N_d$ for binary and symmetry encoding methods, respectively. The results are summarized over 50 independent runs. The table shows the number of CNOT gates (i.e., quantum resources), the number of parameters to optimize, obtainable fidelity, optimization iterations, and classical resources, defined in Eq.~(\ref{eq:classical_resources}), for three different values of \(\phi\).

The results using binary encoding with a hardware-efficient ansatz are shown in the top half of the table, which fails to converge to fidelity \(\mathcal{F} \geq 0.95\) for all values of \(\phi\). More importantly, adding more layers to the ansatz decreases the fidelity, indicating severe trainability degradation. In contrast, the results using symmetry encoding with a total spin preserving ansatz, shown in the bottom half of the table, successfully converge to fidelity \(\mathcal{F} \geq 0.95\) for all tested \(\phi\) values. These results demonstrate the same trade-off seen in the spin-1 benchmark: A lower CNOT count by itself does not guarantee a useful VQE outcome, whereas symmetry encoding improves fidelity and trainability even when it requires additional qubits and, in this spin-\(3/2\) case, larger CNOT counts. Note that these results are presented without implementing any barren plateau mitigation techniques, which could potentially enhance both the performance of the binary and the symmetry encoding methods.


\section{Barren plateau gradient norm diagnostic}
\label{sec:barren_plateau}

Here, we analyze the gradient norm of the spin-1 BBH benchmark. We consider two representative cases: the dimerized phase $\phi = -0.71\pi$, where the cost function is $\langle H\rangle$, and the critical phase $\phi = 0.32\pi$, where the optimization includes an encoding specific penalty term.

For each encoding, depth $L$, and qubit count $N_q$, we sample $R = 500$ independent parameter vectors from $\mathrm{Uniform}[-\pi,\pi]$ and evaluate the exact gradient of the corresponding cost function. In the no-penalty case, the relevant gradient is $g_1$ for both encodings. In the critical case, both encodings use
\begin{equation}\label{eq:bp_grad_total}
  g_{\mathrm{tot}} = g_1 + \beta\, g_2, \qquad \beta = 1,
\end{equation}
where $g_2$ is the gradient of the encoding specific illegitimate state penalty. For symmetry encoding, this penalty is $\mathcal{P}_{i,i+1}$. For binary encoding, it is the analogous binary penalty. Both are defined in the main text. Although the original binary critical phase training uses a larger $\beta$, here we set $\beta = 1$ for both encodings so that the comparison focuses on the encoding and ansatz choice rather than differences in penalty weight. As diagnostic, we use the normalized gradient norm $\|\nabla C\|_2/\sqrt{P}$, where $P$ is the number of variational parameters, and plot the median together with the interquartile range.

\begin{figure}[t]
  \centering
  \begin{subfigure}[b]{0.49\columnwidth}
    \includegraphics[width=\linewidth]{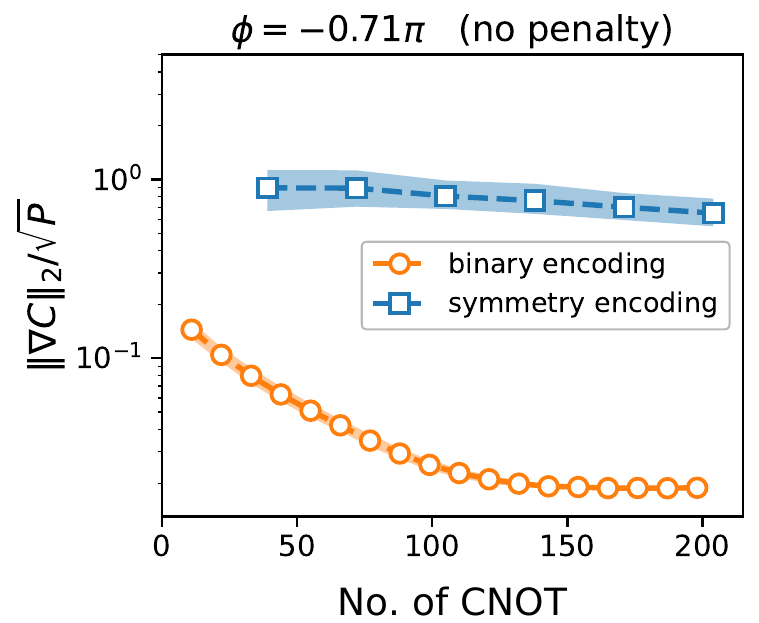}
    \caption{}
    \label{fig:bp_noncrit_cnots}
  \end{subfigure}
  \hfill
  \begin{subfigure}[b]{0.49\columnwidth}
    \includegraphics[width=\linewidth]{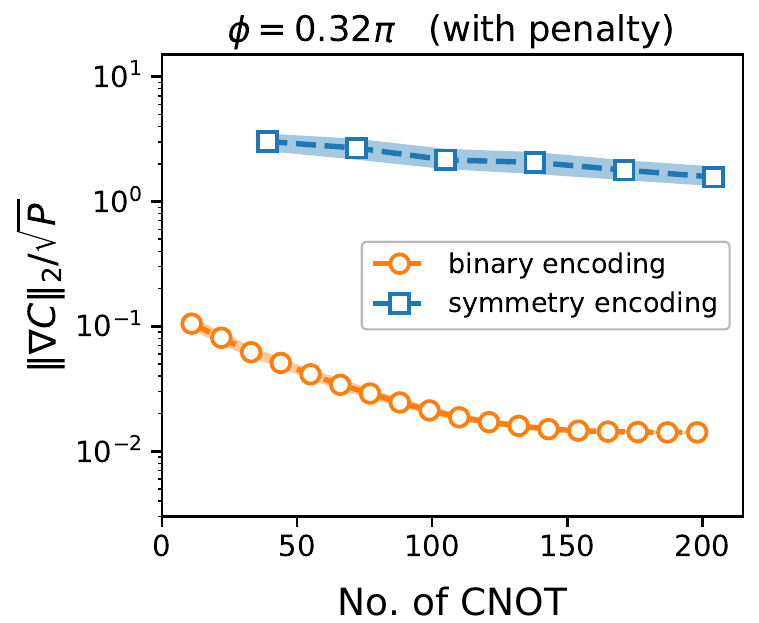}
    \caption{}
    \label{fig:bp_crit_cnots}
  \end{subfigure} \\
  \begin{subfigure}[b]{0.49\columnwidth}
    \includegraphics[width=\linewidth]{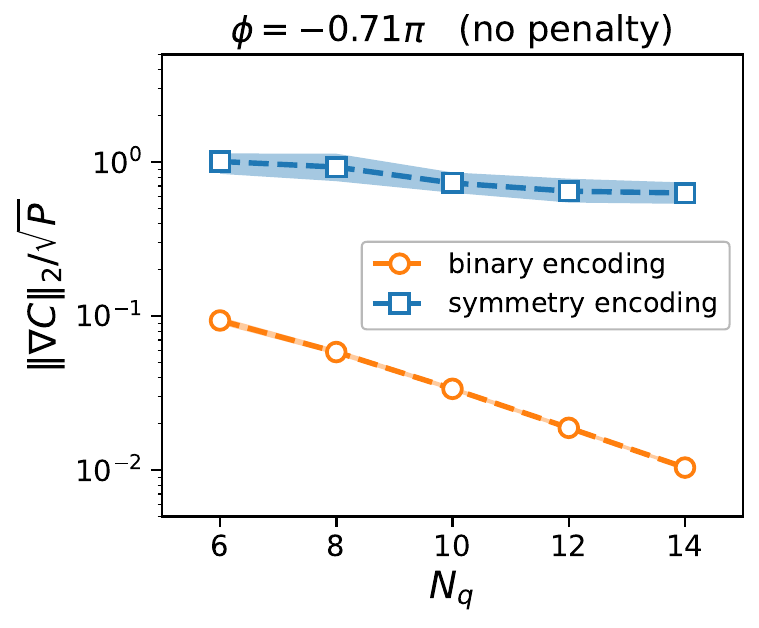}
    \caption{}
    \label{fig:bp_noncrit_N_fixC}
  \end{subfigure}
  \hfill
  \begin{subfigure}[b]{0.49\columnwidth}
    \includegraphics[width=\linewidth]{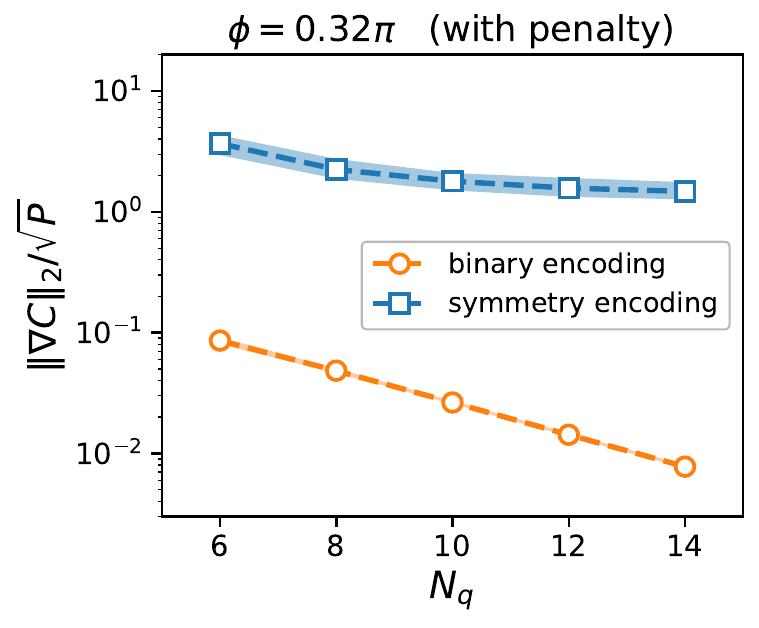}
    \caption{}
    \label{fig:bp_crit_N_fixC}
  \end{subfigure}
  \caption{
    Gradient norm diagnostic for the binary encoding (orange circles) and the symmetry encoding (blue squares). Markers show the median over $R = 500$ random parameter vectors drawn from $\mathrm{Uniform}[-\pi,\pi]$, and the shaded bands show the interquartile range. In the top row, panels (a) and (b) plot the gradient norm vs the total CNOT count at fixed $N_q = 12$. In the bottom row, panels (c) and (d) plot the gradient norm vs qubit count $N_q$ using depths $L_{\mathrm{bin}} = 18$ and $L_{\mathrm{sym}} = 6$, which are chosen deep enough that the gradient norm of each encoding has effectively saturated. The corresponding CNOT counts agree within about $3\%$ at every $N_q$. The left column, panels (a) and (c), corresponds to the dimerized phase $\phi = -0.71\pi$ with cost $\langle H\rangle$. The right column, panels (b) and (d), corresponds to the critical phase $\phi = 0.32\pi$ with encoding specific penalties and $\beta = 1$ for both encodings. In all panels, the gradient norm of the binary encoding decreases much more rapidly than that of the symmetry encoding.
  }
  \label{fig:bp}
\end{figure}

The top row of Fig.~\ref{fig:bp} compares the two encodings at fixed $N_q = 12$ as a function of the total number of CNOT gates. The bottom row sweeps the qubit count over $N_q \in \{6, 8, 10, 12, 14\}$ at per encoding depths $L_{\mathrm{bin}} = 18$ and $L_{\mathrm{sym}} = 6$, both chosen deep enough that the gradient norm for each encoding has already saturated, so the plot reflects system size scaling with depth effects largely removed. The resulting CNOT counts match to within $\sim 3\%$ at every $N_q$, giving a fair comparison. With depth fixed in this saturated regime, the binary gradient norm drops by roughly an order of magnitude between $N_q = 6$ and $N_q = 14$ in both scenarios while the symmetry norm stays flat, and the gap between binary and symmetry widens from $\sim 10\times$ at $N_q = 6$ to $\sim 100\times$ at $N_q = 14$. This is consistent with the barren plateau signature~\cite{McClean2018,Holmes2022} for the binary hardware-efficient ansatz.

\section{Optimizer and noise robustness}

\subsection{Choice of classical optimizer}

\begin{figure}[t]
  \centering
  \begin{subfigure}[b]{0.49\columnwidth}
    \includegraphics[width=\linewidth]{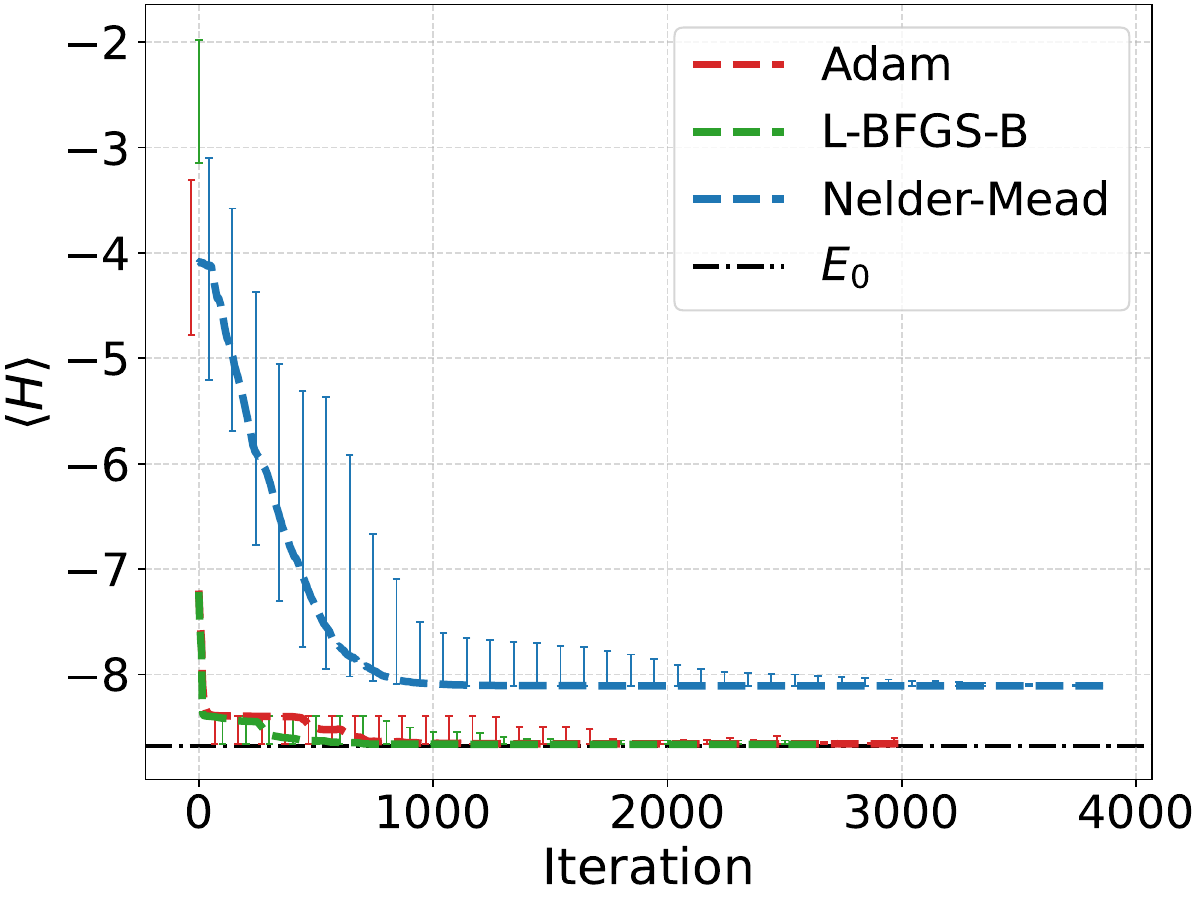}
    \caption{}
    \label{fig:optimizer_a}
  \end{subfigure}
  \hfill
  \begin{subfigure}[b]{0.49\columnwidth}
    \includegraphics[width=\linewidth]{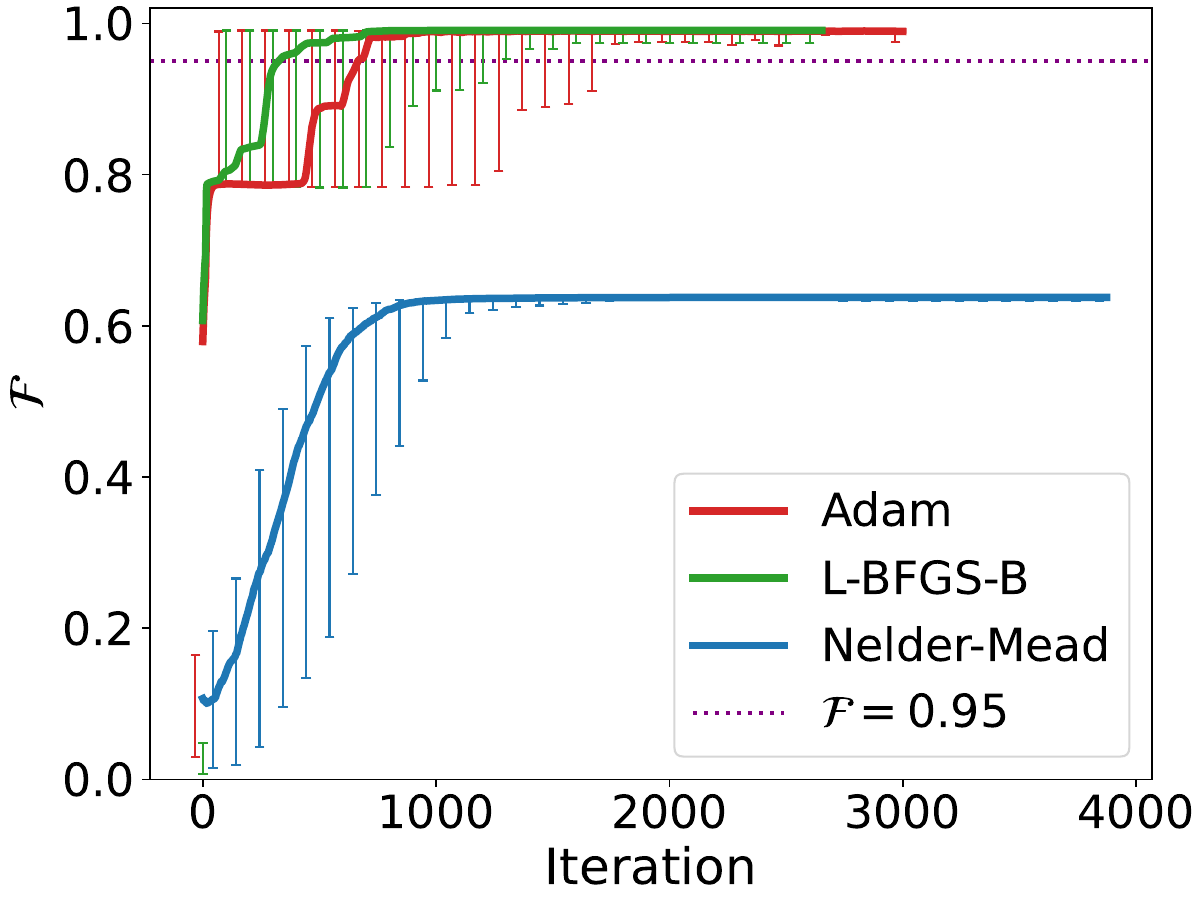}
    \caption{}
    \label{fig:optimizer_b}
  \end{subfigure}
  \caption{
    Comparison of classical optimizers for the spin-1 BBH model at $\phi = -0.71\pi$ with $N_d=6$, using a six-layer symmetry-preserving ansatz. Panel (a) shows the energy expectation value $\langle H \rangle$, and panel (b) shows the median fidelity $\mathcal{F}$ as a function of optimization iteration for Adam (red), L-BFGS-B (green), and Nelder-Mead (blue). The results are summarized over 50 independent runs with random initial parameters.}
  \label{fig:optimizer_res}
\end{figure}

The performance of VQE algorithms can be significantly influenced by the choice of classical optimizer. To investigate this impact, we compare the performance of three distinct optimizers: Adam~\cite{Kingma2014adam}, L-BFGS-B~\cite{Liu1989}, and Nelder-Mead~\cite{Nelder1965simplex}. Adam is a gradient-based optimizer well suited for noisy optimization landscapes, L-BFGS-B is a quasi-Newton method efficient for smooth landscapes, and Nelder-Mead is a gradient-free simplex method. Figure~\ref{fig:optimizer_res} shows the comparison of energy expectation value and fidelity for the spin-1 BBH model in the dimerized phase ($\phi = -0.71\pi$) using a six-layer symmetry-preserving ansatz.

As observed in Fig.~\ref{fig:optimizer_res}, the gradient-free Nelder-Mead optimizer performs significantly worse than Adam and L-BFGS-B for this spin-1 benchmark, failing to reach comparable fidelity over the reported iteration range. More generally, optimizer performance in VQAs depends strongly on the loss landscape and trainability of the ansatz, so we interpret Fig.~\ref{fig:optimizer_res} as evidence for the present system and ansatz rather than as a general necessity of gradient information~\cite{cerezo2021vqa,arrasmith2021gradientfree}. Both Adam and L-BFGS-B effectively converge to the ground state. L-BFGS-B demonstrates faster convergence in terms of optimization iterations compared to Adam. However, it is important to note that L-BFGS-B is computationally more intensive per iteration, as it requires multiple function evaluations for line search in addition to gradient calculations. Adam, while requiring more iterations, performs a fixed number of circuit evaluations per step. Furthermore, Adam is generally more robust against statistical noise, which is critical for implementation on actual quantum hardware. Based on these considerations, we employed the Adam optimizer for the comprehensive simulations presented in this work.

\begin{figure}[t]
  \centering
  \begin{subfigure}[b]{0.49\columnwidth}
    \includegraphics[width=\linewidth]{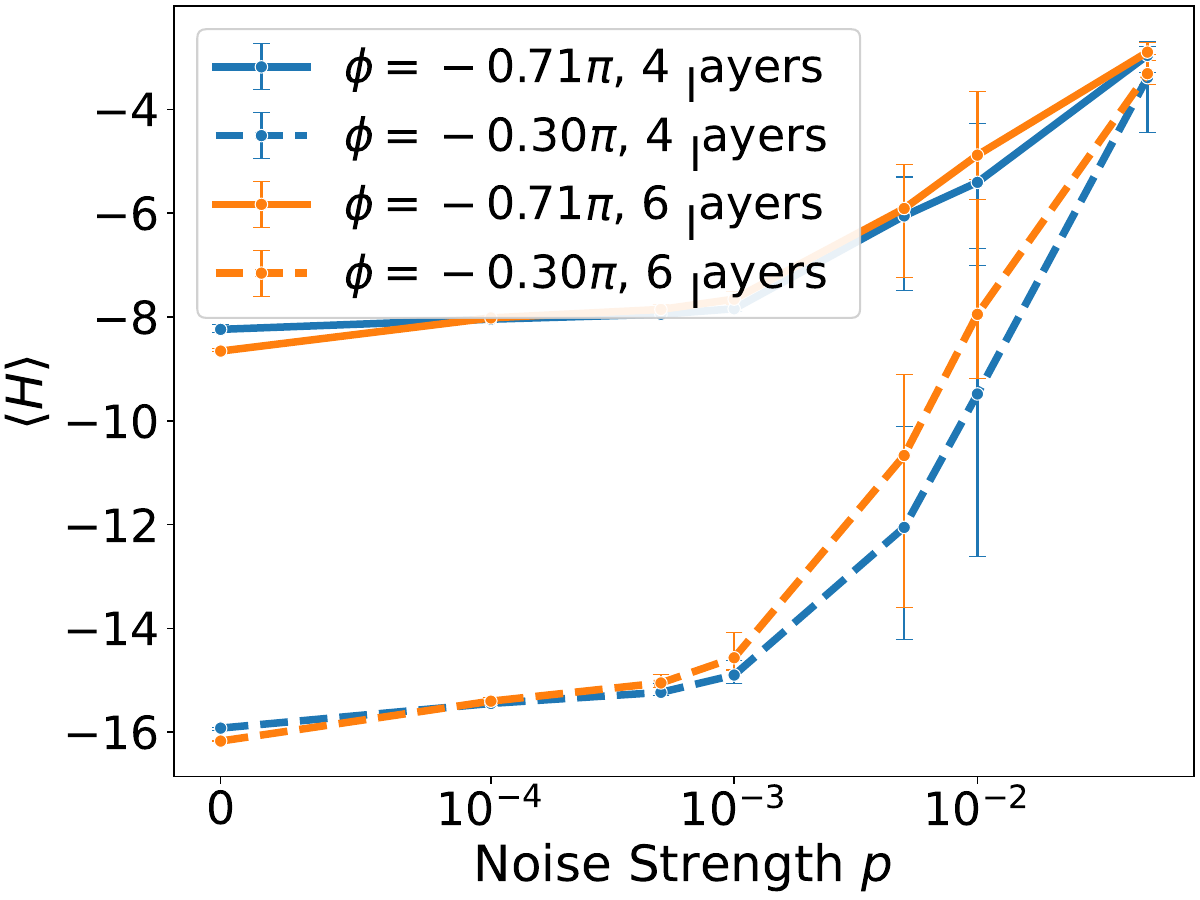}
    \caption{}
    \label{fig:noisy_a}
  \end{subfigure}
  \hfill
  \begin{subfigure}[b]{0.49\columnwidth}
    \includegraphics[width=\linewidth]{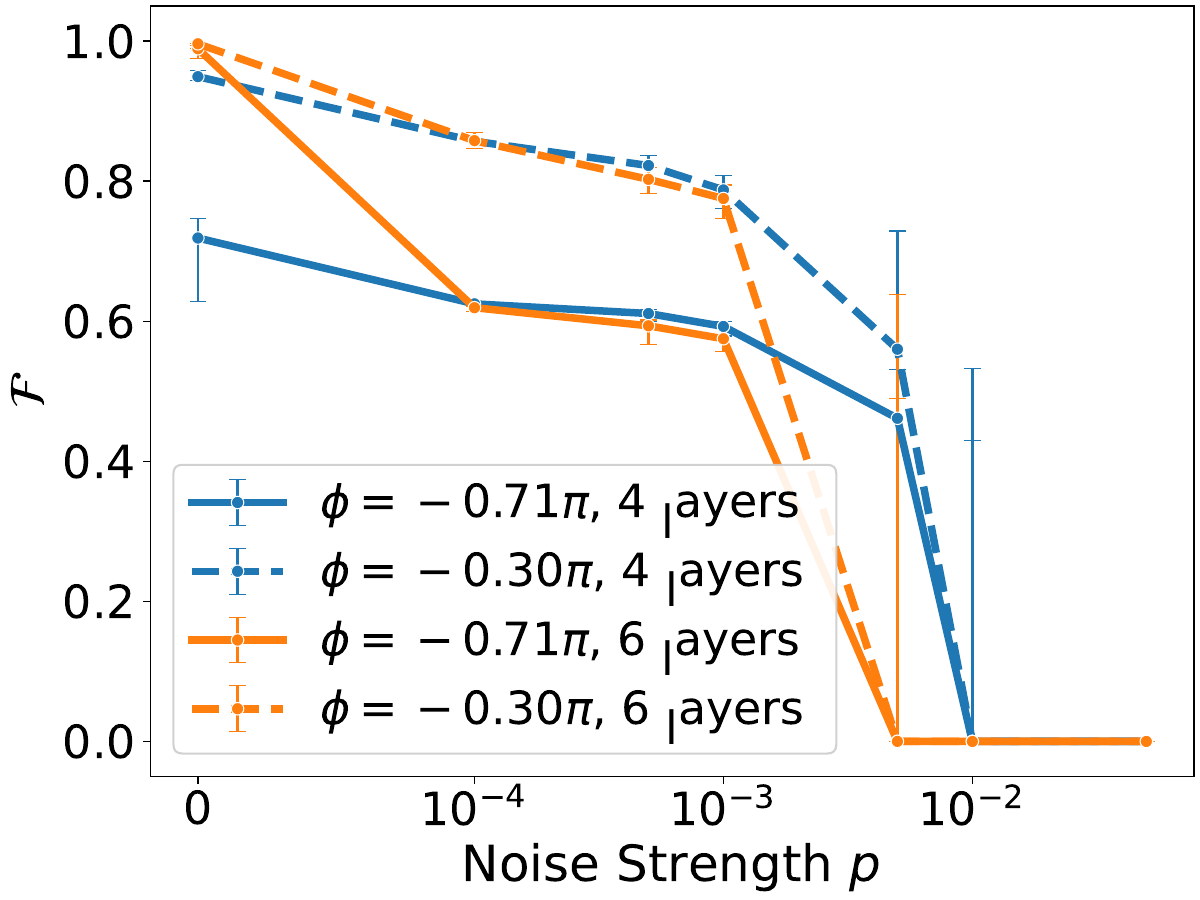}
    \caption{}
    \label{fig:noisy_b}
  \end{subfigure} 
  \caption{
Performance of the VQE ansatz in the presence of depolarizing noise. 
We consider $\phi = -0.71\pi$ and $\phi = -0.30\pi$ in the dimerized phase, and circuit depths with four and six layers.
After every CNOT gate, we apply a two-qubit depolarizing channel with strength $p$, while all single-qubit gates are assumed to be perfect.
Panel (a) shows the energy expectation value $\langle H \rangle$ as a function of $p$, and panel (b) shows the corresponding median fidelity $\mathcal{F}$ with the exact ground state.
Markers indicate the median over 50 independent runs with random initial parameters, and the error bars denote the interquartile range (25th--75th percentile) across runs.}
  \label{fig:noisy_res}
\end{figure}

\subsection{Performance under gate noise}

The results discussed so far were obtained in an ideal, noiseless setting. 
In order to assess how robust the proposed ansatz is to realistic hardware imperfections, we next study the performance of the VQE in the presence of gate noise. 
We adopt a simple but standard noise model in which every two-qubit gate is followed by an identical depolarizing channel, while all single-qubit gates are assumed to be perfect. 
More precisely, after each CNOT gate acting on qubits $i$ and $j$, we apply the two-qubit depolarizing channel
\begin{equation}
\mathcal{D}_p(\rho)
= (1-p)\,\rho + \frac{p}{4} \mathbb{I}_{ij} \otimes \mathrm{Tr}_{ij}(\rho),
\end{equation}
with noise strength $p$, whereas single-qubit rotations are taken to be noiseless. 
The cost function is still evaluated with respect to the exact qubit Hamiltonian, and we retrain the variational parameters for each value of $p$ using the same Adam-based optimization protocol as in the noiseless case.

Figure~\ref{fig:noisy_res} summarizes the dependence of the optimized energy and fidelity on the noise strength. 
We consider two representative points in the dimerized phase, $\phi = -0.71\pi$ and $\phi = -0.30\pi$, and circuit depths with four and six layers.
For each $(\phi,\text{layer})$ and noise strength $p$, we perform 50 independent optimization runs and plot the median final energy $\langle H\rangle$ and fidelity $\mathcal{F}$ with interquartile-range error bars (25th--75th percentiles).

Panel (a) shows that, for small $p$, all configurations converge close to the ground-state energy. 
As $p$ increases, the energy rises monotonically, with the deeper (six-layer) circuits achieving slightly lower energies at very weak noise but degrading faster than the four-layer circuits, reflecting their larger CNOT count.
The fidelities in panel (b) follow the same trend: Deeper circuits perform better in the near-ideal regime, but all configurations eventually suffer a sharp loss of overlap with the exact ground state as noise grows. 
This illustrates the usual trade-off between expressivity and noise robustness: Adding layers improves performance in the noiseless or weak-noise limit, but also makes the ansatz more vulnerable to realistic gate errors.

\section{Discussion} \label{sec:discussion}

We have studied variational simulations of higher-spin systems on qubit-based quantum simulators by comparing binary and symmetry-preserving encodings. Binary encoding minimizes the qubit overhead, requiring only $\lceil \log_2(d) \rceil$ qubits per particle, but it does not naturally preserve the symmetries of the original model in the variational circuit. Symmetry encoding instead uses $d-1$ qubits per particle and supports ansätze that conserve quantities such as $S_{\mathrm{tot}}^2$ and $S_{\mathrm{tot}}^z$. In both cases, the encoded qubit Hilbert space is larger than the physical qudit Hilbert space, which introduces illegitimate states without a qudit counterpart. We showed that these states can be controlled systematically with penalty terms and that, once this issue is addressed, symmetry encoding offers a clear practical advantage in the spin-1 and spin-3/2 benchmarks studied here. In particular, symmetry-preserving circuits reach higher fidelities with lower classical cost and exhibit substantially weaker trainability degradation than the hardware-efficient ansatz under binary encoding, although this advantage does not imply lower quantum resource requirements in terms of qubit number or two-qubit gate count. { The Hamiltonian structure also contributes to this trade-off. For an $m$-body term, the longest Pauli strings act on up to $m\lceil\log_2 d\rceil$ qubits in binary encoding and $m(d-1)$ qubits in symmetry encoding.} These results indicate that symmetry-preserving encodings provide a viable route to simulating a broad class of qudit models on current qubit hardware, although the concrete encoding must still be tailored to the symmetries one wishes to preserve.

\begin{acknowledgments}
A.B. acknowledges support from the National Natural Science Foundation of China
(Grants No. W2541020, No. 12274059, No. 12574528, and No. 1251101297).
\end{acknowledgments}

The project was conceived by A.B. and further developed through discussions with C.L., Z.Z., and M.-H.Y. C.L. developed the numerical codes, and X.X. and Z.Z. optimized their execution. All authors discussed the results. The article was written by A.B. and C.L. with input from all authors.

The authors declare no competing interests.

\section*{Data availability}
The datasets generated and analyzed during the current study are available from the corresponding author upon reasonable request.

\appendix

\section{Numerical details}
\label{app:numerics}

Quantum circuit simulation was performed using Mindquantum~\cite{xu2024mindsporequantum}, utilizing state-vector simulations to calculate the exact expectation values of the Hamiltonian and fidelities. For computing the energy eigenvalues and calculating their corresponding fidelity with the VQE output state, we use the Python toolbox \textsc{Quimb}~\cite{gray2018quimb}. For the primary VQE optimization results, we employed the Adam optimizer. Additionally, for the optimizer comparison analysis, we also utilized the L-BFGS-B and Nelder-Mead algorithms as implemented in \textsc{SciPy}~\cite{2020SciPy-NMeth}.

\newpage
\section{Encoded spin-1 Hamiltonian with symmetry encoding}
\label{app:symmetry_encoding}

The qubit Hamiltonian for the spin-1 BBH model with symmetry encoding, derived using the transformation described, is given by
{\small
\setlength{\jot}{3pt}
\begin{equation}
\begin{split}
\vec{\widetilde{S}}_i \cdot \vec{\widetilde{S}}_{i+1} \mapsto & \frac{1}{4} \Big( \sigma_{j+1}^x \sigma_{j+3}^x + \sigma_{j+1}^x \sigma_{j+2}^x \\
&\quad + \sigma_j^x \sigma_{j+3}^x + \sigma_j^x \sigma_{j+2}^x \\
&\quad + \sigma_{j+1}^y \sigma_{j+3}^y + \sigma_{j+1}^y \sigma_{j+2}^y \\
&\quad + \sigma_j^y \sigma_{j+3}^y + \sigma_j^y \sigma_{j+2}^y \\
&\quad + \sigma_j^z \sigma_{j+2}^z + \sigma_j^z \sigma_{j+3}^z \\
&\quad + \sigma_{j+1}^z \sigma_{j+2}^z + \sigma_{j+1}^z \sigma_{j+3}^z \Big),
\end{split}
\end{equation}
\begin{equation}
\begin{split}
(\vec{\widetilde{S}}_i \cdot \vec{\widetilde{S}}_{i+1})^2 \mapsto & \frac{3}{4} I + \frac{1}{4} \Big( \sigma_{j+2}^x \sigma_{j+3}^x + \sigma_j^x \sigma_{j+1}^x \\
&\quad + \sigma_j^x \sigma_{j+1}^x \sigma_{j+2}^x \sigma_{j+3}^x \Big) \\
&\quad - \frac{1}{8} \Big( \sigma_{j+1}^{z} \sigma_{j+3}^{z} + \sigma_{j+1}^{y} \sigma_{j+3}^{y} \\
&\quad + \sigma_{j+1}^{z} \sigma_{j+2}^{z} + \sigma_{j+1}^{y} \sigma_{j+2}^{y} \\
&\quad + \sigma_j^{z} \sigma_{j+3}^{z} + \sigma_j^{y} \sigma_{j+3}^{y} \\
&\quad + \sigma_j^{z} \sigma_{j+2}^{z} + \sigma_j^{y} \sigma_{j+2}^{y} \\
&\quad + \sigma_{j+1}^x \sigma_{j+3}^x + \sigma_{j+1}^x \sigma_{j+2}^x \\
&\quad + \sigma_j^x \sigma_{j+3}^x + \sigma_j^x \sigma_{j+2}^x \Big) \\
&\quad + \frac{1}{8} \Big( \sigma_j^y \sigma_{j+1}^x \sigma_{j+2}^y \sigma_{j+3}^x \\
&\quad + \sigma_j^z \sigma_{j+1}^x \sigma_{j+2}^z \sigma_{j+3}^x \\
&\quad + \sigma_j^y \sigma_{j+1}^x \sigma_{j+2}^x \sigma_{j+3}^y \\
&\quad + \sigma_j^z \sigma_{j+1}^x \sigma_{j+2}^x \sigma_{j+3}^z \\
&\quad + \sigma_j^x \sigma_{j+1}^y \sigma_{j+2}^y \sigma_{j+3}^x \\
&\quad + \sigma_j^x \sigma_{j+1}^z \sigma_{j+2}^z \sigma_{j+3}^x \\
&\quad + \sigma_j^x \sigma_{j+1}^y \sigma_{j+2}^x \sigma_{j+3}^y \\
&\quad + \sigma_j^x \sigma_{j+1}^z \sigma_{j+2}^x \sigma_{j+3}^z \\
&\quad + \sigma_j^z \sigma_{j+1}^y \sigma_{j+2}^z \sigma_{j+3}^y \\
&\quad + \sigma_j^z \sigma_{j+1}^y \sigma_{j+2}^y \sigma_{j+3}^z \\
&\quad + \sigma_j^y \sigma_{j+1}^z \sigma_{j+2}^z \sigma_{j+3}^y \\
&\quad + \sigma_j^y \sigma_{j+1}^z \sigma_{j+2}^y \sigma_{j+3}^z \Big) \\
&\quad + \frac{1}{4} \Big( \sigma_{j+2}^y \sigma_{j+3}^y + \sigma_j^y \sigma_{j+1}^y \\
&\quad + \sigma_j^y \sigma_{j+1}^y \sigma_{j+2}^y \sigma_{j+3}^y \\
&\quad + \sigma_{j+2}^z \sigma_{j+3}^z + \sigma_j^z \sigma_{j+1}^z \\
&\quad + \sigma_j^z \sigma_{j+1}^z \sigma_{j+2}^z \sigma_{j+3}^z \Big).
\end{split}
\end{equation}
}

\newpage
\section{Encoded spin-1 Hamiltonian with binary encoding}
\label{app:binary_encoding}

Here, we provide the exact form of result obtained by applying the binary encoding method to the spin-1 BBH Hamiltonian, as shown in Eqs.~\eqref{eq:binary_mapping} and \eqref{eq:binary_transformation}. The qubit Hamiltonian for the spin-1 BBH model, after implementing the transformation described in Eq.~\eqref{eq:op_mapping}, is as follows:

{\small
\setlength{\jot}{3pt}
\begin{equation}
\begin{split}
\vec{\widetilde{S}}_i \cdot \vec{\widetilde{S}}_{i+1} \mapsto & \frac{1}{8} \big( \sigma_{j+1}^x \sigma_{j+3}^x + \sigma_{j+1}^x \sigma_{j+2}^z \sigma_{j+3}^x \\
  &\quad + \sigma_{j+1}^x \sigma_{j+2}^x - \sigma_{j+1}^x \sigma_{j+2}^x \sigma_{j+3}^z \\
  &\quad + \sigma_j^z \sigma_{j+1}^x \sigma_{j+3}^x + \sigma_j^z \sigma_{j+1}^x \sigma_{j+2}^z \sigma_{j+3}^x \\
  &\quad + \sigma_j^z \sigma_{j+1}^x \sigma_{j+2}^x - \sigma_j^z \sigma_{j+1}^x \sigma_{j+2}^x \sigma_{j+3}^z \\
  &\quad + \sigma_j^x \sigma_{j+3}^x + \sigma_j^x \sigma_{j+2}^z \sigma_{j+3}^x \\
  &\quad + \sigma_j^x \sigma_{j+2}^x - \sigma_j^x \sigma_{j+2}^x \sigma_{j+3}^z \\
  &\quad - \sigma_j^x \sigma_{j+1}^z \sigma_{j+3}^x - \sigma_j^x \sigma_{j+1}^z \sigma_{j+2}^z \sigma_{j+3}^x \\
  &\quad - \sigma_j^x \sigma_{j+1}^z \sigma_{j+2}^x + \sigma_j^x \sigma_{j+1}^z \sigma_{j+2}^x \sigma_{j+3}^z \\
  &\quad + \sigma_{j+1}^y \sigma_{j+3}^y + \sigma_{j+1}^y \sigma_{j+2}^z \sigma_{j+3}^y \\
  &\quad + \sigma_{j+1}^y \sigma_{j+2}^y - \sigma_{j+1}^y \sigma_{j+2}^y \sigma_{j+3}^z \\
  &\quad + \sigma_j^z \sigma_{j+1}^y \sigma_{j+3}^y + \sigma_j^z \sigma_{j+1}^y \sigma_{j+2}^z \sigma_{j+3}^y \\
  &\quad + \sigma_j^z \sigma_{j+1}^y \sigma_{j+2}^y - \sigma_j^z \sigma_{j+1}^y \sigma_{j+2}^y \sigma_{j+3}^z \\
  &\quad + \sigma_j^y \sigma_{j+3}^y + \sigma_j^y \sigma_{j+2}^z \sigma_{j+3}^y \\
  &\quad + \sigma_j^y \sigma_{j+2}^y - \sigma_j^y \sigma_{j+2}^y \sigma_{j+3}^z \\
  &\quad - \sigma_j^y \sigma_{j+1}^z \sigma_{j+3}^y - \sigma_j^y \sigma_{j+1}^z \sigma_{j+2}^z \sigma_{j+3}^y \\
  &\quad - \sigma_j^y \sigma_{j+1}^z \sigma_{j+2}^y + \sigma_j^y \sigma_{j+1}^z \sigma_{j+2}^y \sigma_{j+3}^z \big) \\
  &\quad + \frac{1}{4} \big( \sigma_j^z \sigma_{j+2}^z + \sigma_j^z \sigma_{j+3}^z \\
  &\quad + \sigma_{j+1}^z \sigma_{j+2}^z + \sigma_{j+1}^z \sigma_{j+3}^z \big),
\end{split}
\end{equation}
\begin{equation}
\begin{split}
(\vec{\widetilde{S}}_i \cdot \vec{\widetilde{S}}_{i+1})^2 \mapsto & \frac{3}{4} I + \frac{1}{4} \Big(\sigma_{j+2}^z + \sigma_j^z - \sigma_{j+3}^z \\
  &\quad - \sigma_j^z \sigma_{j+3}^z - \sigma_{j+1}^z \sigma_{j+2}^z - \sigma_{j+1}^z\Big) \\
  &\quad + \frac{1}{8} \Big(\sigma_j^x \sigma_{j+1}^x \sigma_{j+2}^x \sigma_{j+3}^x + \sigma_j^y \sigma_{j+1}^x \sigma_{j+2}^y \sigma_{j+3}^x \\
  &\quad - \sigma_j^x \sigma_{j+1}^x \sigma_{j+2}^y \sigma_{j+3}^y + \sigma_j^y \sigma_{j+1}^x \sigma_{j+2}^x \sigma_{j+3}^y \\
  &\quad + \sigma_j^z \sigma_{j+1}^x \sigma_{j+2}^x \sigma_{j+3}^z + \sigma_j^x \sigma_{j+1}^y \sigma_{j+2}^y \sigma_{j+3}^x \\
  &\quad + \sigma_j^y \sigma_{j+1}^y \sigma_{j+2}^y \sigma_{j+3}^y + \sigma_j^x \sigma_{j+1}^y \sigma_{j+2}^x \sigma_{j+3}^y \\
  &\quad + \sigma_j^x \sigma_{j+1}^x \sigma_{j+2}^x \sigma_{j+3}^z - \sigma_j^z \sigma_{j+1}^y \sigma_{j+2}^y \\
  &\quad - \sigma_{j+1}^x \sigma_{j+2}^x + \sigma_j^z \sigma_{j+1}^y \sigma_{j+2}^y \sigma_{j+3}^z \\
  &\quad - \sigma_j^y \sigma_{j+3}^y + \sigma_j^x \sigma_{j+1}^z \sigma_{j+2}^z \sigma_{j+3}^x \\
  &\quad - \sigma_j^y \sigma_{j+2}^z \sigma_{j+3}^y + \sigma_j^x \sigma_{j+1}^z \sigma_{j+3}^x \\
  &\quad + \sigma_j^y \sigma_{j+1}^z \sigma_{j+3}^y - \sigma_j^x \sigma_{j+2}^z \sigma_{j+3}^x \\
  &\quad + \sigma_j^y \sigma_{j+1}^z \sigma_{j+2}^z \sigma_{j+3}^y - \sigma_j^x \sigma_{j+3}^x \Big) \\
  &\quad + \frac{1}{4} \Big(\sigma_{j+2}^z \sigma_{j+3}^z + \sigma_j^z \sigma_{j+1}^z \\
  &\quad + \sigma_j^z \sigma_{j+1}^z \sigma_{j+2}^z \sigma_{j+3}^z\Big).
\end{split}
\end{equation}
}
\clearpage

\bibliographystyle{apsrev4-2-author-truncate}
\bibliography{References}

\end{document}